\documentclass[english]{IEEEtran}
\usepackage[T1]{fontenc}
\usepackage{xcolor}
\usepackage{float}
\usepackage{amsmath}
\usepackage{amsthm}
\usepackage{amssymb}
\usepackage{stmaryrd}
\usepackage{graphicx}

\floatstyle{ruled}
\newfloat{algorithm}{tbp}{loa}
\providecommand{\algorithmname}{Algorithm}
\floatname{algorithm}{\protect\algorithmname}

\newtheorem{theorem}{Theorem}
\newtheorem{lemma}{Lemma}

\newtheorem{proposition}{Proposition}

\usepackage{amsmath}
\usepackage{amssymb}
\usepackage{graphicx}
\usepackage{colortbl}

\author{\IEEEauthorblockN{Bowen Li, Haotian Zhang, Mu Jia, Junting Chen, Nikolaos Pappas}

\thanks{Junting Chen is Corresponding author.}
\thanks{Bowen Li was with the School of Science and Engineering (SSE), 
and the Shenzhen Future Network of Intelligence Institute (FNii--Shenzhen), The Chinese University of Hong Kong, Shenzhen, Guangdong 518172, China.
He is now with Department of Computer and Information Science, Link{\"o}ping University, 58183, Link{\"o}ping, Sweden. (email: bowen.li@liu.se).}
\thanks{Haotian Zhang, Mu Jia, and Junting Chen are with the School of Science and Engineering (SSE), 
and the Shenzhen Future Network of Intelligence Institute (FNii--Shenzhen), The Chinese University of Hong Kong, Shenzhen, Guangdong 518172, China.
(email: haotianzhang@link.cuhk.edu.cn, mujia1@link.cuhk.edu.cn, and juntingc@cuhk.edu.cn).}
\thanks{Nikolaos Pappas is with Department of Computer and Information Science, Link{\"o}ping University, 58183, Link{\"o}ping, Sweden. (email: nikolaos.pappas@liu.se).}

}

\usepackage[acronym]{glossaries}
\newcommand{\newac}{\newacronym}
\newcommand{\ac}{\gls}
\newcommand{\Ac}{\Gls}
\newcommand{\acpl}{\glspl}

\newac{speb}{SPEB}{square position error bound}
\newac[plural=EFIMs,firstplural=Fisher information matrices (EFIMs)]{efim}{EFIM}{Fisher information matrix}
\newac{ne}{NE}{Nash equilibrium}
\newac{mse}{MSE}{mean squared error}
\newac{toa}{TOA}{time-of-arrival}
\newac{snr}{SNR}{signal-to-noise ratio}
\newac{lan}{LAN}{local area network}
\newac{psd}{PSD}{positive semidefinite}
\newac{pd}{PD}{positive definite}
\newac{wrt}{w.r.t.}{with respect to}
\newac{lhs}{L.H.S.}{left hand side}
\newac{wp1}{w.p.1}{with probability 1}
\newac{kkt}{KKT}{Karush-Kuhn-Tucker}
\newac{wlog}{w.l.o.g.}{without loss of generality}
\newac{mle}{MLE}{maximum likelihood estimation}
\newac{gps}{GPS}{global positioning system}
\newac{rssi}{RSSI}{received signal strength indicator}
\newac{mimo}{MIMO}{multiple-input multiple-output}
\newac{csi}{CSI}{channel state information}
\newac{fdd}{FDD}{frequency division duplexing}
\newac{ms}{MS}{mobile station}
\newac{bs}{BS}{base station}
\newac{d2d}{D2D}{device-to-device}
\newac{slnr}{SLNR}{signal-to-interference-leakage-and-noise-ratio}
\newac{ula}{ULA}{uniform linear antenna array}
\newac{pas}{PAS}{power angular spectrum}
\newac{mmse}{MMSE}{minimum mean square error}
\newac{zf}{ZF}{zero-forcing}
\newac{rzf}{RZF}{regularized zero-forcing}
\newac{as}{AS}{angular spread}
\newac{aod}{AOD}{angle of departure}
\newac{iid}{i.i.d.}{independent and identically distributed} 
\newac{sinr}{SINR}{signal-to-interference-and-noise ratio}
\newac{tdd}{TDD}{time-division duplex}
\newac{rvq}{RVQ}{random vector quantization}
\newac{rhs}{R.H.S.}{right hand side}
\newac{mrc}{MRC}{maximum ratio combining}
\newac{cdf}{CDF}{cumulative distribution function}
\newac{a.s.}{a.s.}{almost surely}
\newac{los}{LOS}{line-of-sight}
\newac{jsdm}{JSDM}{joint spatial division and multiplexing}
\newac{map}{MAP}{maximum a posteriori}
\newac{klt}{KLT}{Karhunen-Lo\`eve Transform}
\newac{lbe}{LBE}{link bargaining equilibrium}
\newac{se}{SE}{Stackelberg equilibrium}
\newac{uav}{UAV}{unmanned aerial vehicle}
\newac{nlos}{NLOS}{non-line-of-sight}
\newac{pdf}{PDF}{probability density function}
\newac{em}{EM}{expectation-maximization}
\newac{knn}{KNN}{$k$-nearest neighbor}
\newac{svd}{SVD}{singular value decomposition}
\newac{nmf}{NMF}{non-negative matrix factorization}
\newac{umf}{UMF}{unimodality-constrained matrix factorization}
\newac{rmse}{RMSE}{rooted mean squared error}
\newac{olos}{OLOS}{obstructed line-of-sight}
\newac{mmw}{mmW}{millimeter wave}
\newac{ber}{BER}{bit error rate}
\newac{rss}{RSS}{received signal strength}
\newac{lp}{LP}{linear program}
\newac{ufw}{U-FW}{unimodal Frank-Wolfe}
\newac{utf}{UTF}{unimodality-constrained tensor factorization}
\newac{fw}{FW}{Frank-Wolfe}
\newac{iot}{IoT}{Internet-of-Things}
\newac{mae}{MAE}{mean absolute error}
\newac{crb}{CRB}{Cram\'er-Rao bound}
\newac{aoa}{AoA}{angle of arrival}
\newac{wcl}{WCL}{weighted centroid localization}

\newac{psd2}{PSD}{power spectral density}
\newac{slf}{SLF}{spatial loss function}
\newac{btd}{BTD}{block term tensor decomposition}
\newac{tps}{TPS}{thin plate spline}
\newac{nmse}{NMSE}{normalized mean squared error}
\newac{cp}{CP}{cyclic prefix}
\newac{cfo}{CFO}{carrier frequency offset}
\newac{sic}{SIC}{successive interference cancellation}
\newac{ml}{ML}{maximum likelihood}
\newac{crlb}{CRLB}{Cramer-Rao lower bound}
\newac{nr}{NR}{new radio}
\newac{ss}{SS}{synchronization signal}
\newac{fim}{FIM}{Fisher information matrix}
\newac{ofdm}{OFDM}{orthogonal frequency division multiplexing}
\newac{ris}{RIS}{reconfigurable intelligent surfaces}
\newac{ssb}{SSB}{synchronization signal block}
\newac{glrt}{GLRT}{generalized likelihood ratio test}
\newac{ue}{UE}{user equipment}
\newac{to}{TO}{timing offset}
\newac{pss}{PSS}{primary synchronization signal}
\newac{sss}{SSS}{secondary synchronization signal}
\newac{pbch}{PBCH}{physical broadcast channel}
\newac{pci}{PCI}{physical cell identity}
\newac{dmrs}{DMRS}{demodulation reference signal}
\newac{ifft}{IFFT}{inverse fast Fourier transform}
\newac{fft}{FFT}{fast Fourier transform}
\newac{rb}{RB}{resource block}
\newac{re}{RE}{resource element}
\newac{mib}{MIB}{master information block}
\newac{usrp}{USRP}{universal software radio peripheral}
\newac{wss}{WSS}{wide-sense stationary}
\newac{doa}{DOA}{direction of arrival}

\usepackage[noadjust]{cite}
\usepackage{color}  
\usepackage{graphicx,psfrag,cite,subfigure}

\begin{document}
\title{Joint CFO-Channel Estimation under\\ Strong Inter-Cell Interference
for\\ Low-Altitude Radio Mapping}
\maketitle
\begin{abstract}
Extending terrestrial networks into low-altitude airspace provides a practical approach for supporting aerial services, where reliable
Low-altitude radio maps are essential for characterizing terrestrial \acpl{bs} coverage and guiding system design. This paper investigates per-cell per-beam radio mapping in dense multi-cell low-altitude environments, where conventional strongest-link-oriented processing is insufficient for comprehensive radio-environment awareness and becomes ineffective under severe co-channel interference. This paper proposes a multi-block joint \ac{cfo}-channel estimation algorithm. Specifically, we introduce a kinematic coherence condition that justifies a frame-constant-\ac{cfo} and block-constant-amplitude signal model, on which a multi-block estimator is developed to combine multiple coherent blocks. \Ac{crlb} analysis shows that the estimation accuracy is governed by the aggregated \ac{sinr} across coherent blocks, rather than by any individual block, thereby enabling reliable weak-signal estimation.
Moreover, the accuracy improves as the number of synchronization-signal blocks per frame increases and the motion-induced variation decreases. This indicates that the proposed method can naturally benefit from future systems with denser spatial-beam deployments, while also guiding trajectory design of aerial measurement platforms to preserve sampling coherence. Simulations show that the proposed framework achieves high detection and estimation accuracy under strong interference, even at \ac{sinr} levels as low as $-30$ dB. Field tests conducted at an altitude of $150$ m further demonstrate its capability for per-cell per-beam radio mapping of more than ten overlapping \acpl{bs}.
\end{abstract}


\section{Introduction\glsresetall\label{sec:intro}}

The low-altitude economy, encompassing activities such as automated logistics and real-time infrastructure inspection within the airspace from ground level up to 3000 meters, has expanded rapidly, creating a critical demand for low-altitude wireless communications~\cite{SonLinWanSun:M25,JiaLiZhuLi:M25}. Extending existing terrestrial wireless networks to low-altitude airspace
offers the most pragmatic solution. This is primarily because: i) terrestrial wireless infrastructure is already densely deployed, providing a widespread and readily available infrastructure; ii) modern \acpl{bs} employ massive antenna arrays, which offer significant potential for spatial multiplexing and precise 3D beamforming; and iii) the low-altitude environment features largely unobstructed propagation paths, leading to favorable \ac{los} conditions.

However, antennas at terrestrial \acpl{bs} are primarily optimized for ground users, with their main lobes intentionally tilted downwards and only unintended, irregular side lobes radiating towards the sky~\cite{YanDinMaoLin:J19,MdIsaWalAru:J21,GerGarAzaLoz:J24}. Consequently, low-altitude coverage arises from a complex superposition of direct signals from these side lobes and multi-path signals from main lobes reflecting off the ground. This creates a highly heterogeneous signal landscape that defies conventional, monotonic decay models, making it extremely difficult to predict. Therefore, to effectively utilize terrestrial networks for aerial services, such as predictive communications~\cite{CheLiSunCui:M26}, flight trajectories design~\cite{ChaLau:J22}, or \ac{bs} beams optimization for aerial users~\cite{ZhoZenLiYan:A25}, it is first imperative to chart the actual radio environment. This establishes an urgent need for accurate {\em low-altitude radio maps} to truly understand the coverage provided by terrestrial infrastructure in the low-altitude airspace.

State-of-the-art radio mapping~\cite{RomKim:M22} largely focuses on data-level techniques, such as matrix completion~\cite{Sunche:J24}
and learning-based prediction~\cite{ShrFuHon:J22}, or model-driven approaches like digital twins~\cite{WanZhaNieYu:M25} and ray-tracing~\cite{HeAiGuaWan:J19}. The former class of methods infers a complete map from a sparse set of trusted measurements, while the latter simulates the radio environment based on detailed geographical models. Our focus, however, is not at the data or model level, but at the physical layer, where we tackle the fundamental challenge of constructing the trusted, ground-truth measurements from raw synchronization signals emitted by terrestrial wireless infrastructure, thereby providing the foundational technology upon which both data- and model-driven techniques depend. Using the 5G \ac{nr} framework as an example, the \ac{ss} burst inherently supports per-cell and per-beam radio mapping by providing requisite spatial identifiers alongside dedicated sequences for precise \ac{cfo} and channel gain estimation~\cite{3gpp:ts38331,3gpp:ts38213,3gpp:ts38211}. Detailed signal structures are deferred to Appendix~\ref{sec:5GNR}.

\begin{figure}
\centering{}\includegraphics[width=0.9\columnwidth]{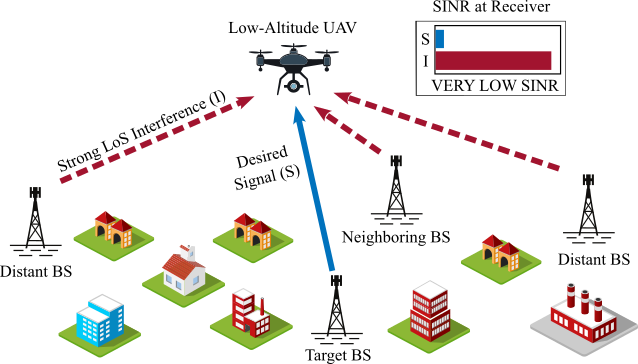}\caption{\label{fig:low_alritude_strong_interference_illu}Low-altitude radio mapping suffers from severe interference, as signals from over 10 \acpl{bs} may be aggregated at the receiver due to LOS propagation; furthermore, \ac{cfo} may degrade the orthogonality of synchronization signals.}
\end{figure}

However, as illustrated in Fig.~\ref{fig:low_alritude_strong_interference_illu}, practical low-altitude radio map construction is fundamentally impeded by severe co-channel interference. Aerial receivers frequently maintain simultaneous \ac{los} to numerous terrestrial transmitters, causing disparate synchronization signals to arrive with comparable powers and partial time-frequency overlap. This pervasive interference makes conventional estimation methods inadequate. Existing algorithms primarily
isolate the strongest signal for link establishment ~\cite{TunRivGarMel:J23,ZhoCheYanChe:J24,YonSawNagSuy:J25,DonCheJuZho:J25,MathWorks:CellSearch}, but aerial mapping requires reliable estimation of both dominant and weak signals. As detailed in Appendix~\ref{sec:5GNR}, such interference disrupts initial delay, cell, and beam detection, precipitating cascaded errors in subsequent \ac{cfo} and channel estimation. Furthermore, aerial receivers are typically mounted on high-mobility platforms (\emph{e.g.}, \acpl{uav}), introducing severe kinematic impairments such as rapid \ac{cfo} drift. Therefore, realizing high-fidelity low-altitude radio mapping necessitates formulating a rigorous kinematic channel model coupled with a robust multi-source framework, capable of resolving superimposed synchronization signals to extract reliable per-cell and per-beam measurements.

\Ac{sic}~\cite{MorKuoPun:J07,ParChu:J12} can, in principle, mitigate compounded interference, under strong interference conditions, but
even small errors in the dominant signal's \ac{cfo} and channel estimates leave residuals whose power remains higher than that of the weaker signals, rendering them effectively undetectable and yielding severely biased estimates. Early works exploited the \ac{cp} of the \ac{ofdm} symbols or dedicated synchronization sequences to jointly estimate the channel and \ac{cfo} in single-transmitter systems~\cite{VanSanBor:J97,LvHuaJie:J05}. Subsequent studies have extended these frameworks to multi-transmitter architectures, such as multi-point relay systems~\cite{NasMehBloDur:J12}, coordinated multi-cell systems~\cite{GesHanHuaSha:J10}, and distributed antenna systems~\cite{XuLarJorLi:J25}, and design joint estimation algorithms to cope with severe co-channel interference and multiple
frequency offsets~\cite{ZarCav:J08,WanXiaYin:J08,TsaHuaCheYan:J13,SalNasMehXia:J17,ZhaGaoJinLin:J18,KunUnnSarLar:J24,SinKumMajSat:J25}.
To summarize, existing approaches for joint \ac{cfo}-channel estimation may only work in the scenario with a small number of weak interferers ({\em e.g.}, 3-4 interferers). For an extremely low \ac{sinr}, a typical case in low-altitude radio map surveying, the conventional \ac{cfo} estimation may fail, and the corresponding channel estimation may collapse.

In this paper, we study the joint \ac{cfo}-channel estimation problem under strong inter-cell interference for per-cell and per-beam low-altitude radio mapping. By analyzing the coherence structure among synchronization signals, we develop a multi-block processing framework for joint estimation, which significantly improves weak-signal detection and parameter estimation. Specifically, this paper makes the following contributions.
\begin{itemize}
\item We characterize the effect of platform kinematics on synchronization waveforms and establish that, under a prescribed kinematic budget, the channel amplitude remains approximately constant within each synchronization block, while the \ac{cfo} remains approximately constant across synchronization blocks within one synchronization frame. Based on these properties, we build an \ac{ml}-based multi-block algorithm for joint \ac{cfo} and channel estimation under strong inter-cell interference. We further show by \ac{crlb} analysis that the estimation error variance depends on the aggregated \ac{sinr} across coherent blocks, and decreases with the number of blocks and with a smaller consumed kinematic budget.
\item We numerically show that the proposed method can reliably detect, estimate, and reconstruct synchronization signals at \ac{sinr} as low as $-30\text{ dB}$ under unknown \ac{cfo}. Our field tests further demonstrate the construction of per-beam coverage maps for more than $10$ overlapping \acpl{bs} at an altitude of $150$ meters. It is found that despite strong received power ($\ge-65\text{ dBm}$) across nearly the entire flight area, the measured \ac{sinr} rarely exceeds $10\text{ dB}$, highlighting the need for careful interference management in low-altitude airspace.
\end{itemize}

The rest of the paper is organized as follows. Section~\ref{sec:System-Model} builds the system model. Section~\ref{sec:Kinematic_analysis} analyzes the kinematic condition for \ac{cfo} coherence.
Section~\ref{sec:Estimation} develops the multi-block processing for joint \ac{cfo} and channel estimation. Section~\ref{sec:Performance-Guarantee} presents the theoretical performance analysis.
Numerical and experimental results are presented in Section~\ref{sec:evaluation}, and conclusions are given in Section~\ref{sec:conclusion}.

\section{System Model\label{sec:System-Model}}

This section establishes the signal model for synchronization-based channel and \ac{cfo} estimation. We first build a short-term constant-amplitude, linear-phase channel representation, and then formulate the synchronization-frame-based multi-block received-signal model for joint parameter estimation.

\subsection{\label{subsec:Channel_model}Channel Model}

We model the time-varying complex channel as $h\left(t\right)=\sqrt{\beta\left(t\right)}\tilde{h}\left(t\right)$,
where $\beta\left(t\right)$ represents the large-scale fading and
$\tilde{h}\left(t\right)$ represents the normalized small-scale Rician
fading. Specifically,
\begin{equation}
\tilde{h}\left(t\right)=\sqrt{\frac{\kappa}{\kappa+1}}e^{j2\pi ft}+\sqrt{\frac{1}{\kappa+1}}h_{\text{d}}\left(t\right)\label{eq:def_rician}
\end{equation}
where $\kappa$ is the Rician factor, $f$ is the effective frequency
offset of the dominant component, and $h_{\text{d}}\left(t\right)$
is the diffuse multipath component, modeled as a zero-mean \ac{wss}
complex Gaussian process with unit variance \cite{TseVis:B05}.

To characterize the short-term channel evolution, define the normalized
temporal autocorrelation at time lag $\delta_{t}$ as $\rho_{h}\left(\delta_{t}\right)\triangleq\mathbb{E}\left\{ h\left(t\right)h^{*}\left(t+\delta_{t}\right)\right\} /\mathbb{E}\{|h(t)|^{2}\}$.
For the considered 5G \ac{nr} synchronization structure, the interval
of interest is on the order of one \ac{ssb}, with duration approximately
0.1 ms, as described in Appendix~\ref{sec:5GNR}. Since the large-scale
fading varies negligibly over this interval, and under the Clarke/Jakes
model \cite{Cla:J68,WilCox:B94}, $\rho_{h}\left(\delta_{t}\right)$
can be approximated by $\frac{\kappa}{\kappa+1}e^{-j2\pi f\delta_{t}}+\frac{1}{\kappa+1}J_{0}\left(2\pi\bar{f}_{d}\delta_{t}\right)$,
where $J_{0}\left(\cdot\right)$ is the zeroth-order Bessel function
of the first kind and $\bar{f}_{d}$ is the maximum Doppler frequency.
Over such a short interval, \emph{i.e.}, $\delta_{t}\approx0.1$ ms,
$J_{0}\left(2\pi f_{m}\delta_{t}\right)\approx1$, so $|\rho_{h}(\delta_{t})|\approx1$
and $\angle\rho_{h}(\delta_{t})\approx-\frac{\kappa}{\kappa+1}2\pi f\delta_{t}$.
Accordingly, over the interval of interest, the channel amplitude
can be approximated as quasi-static, while the phase is locally linear
in $\delta_{t}$ and mainly governed by the dominant frequency offset.

\begin{figure}
\centering{}\includegraphics[width=1\columnwidth]{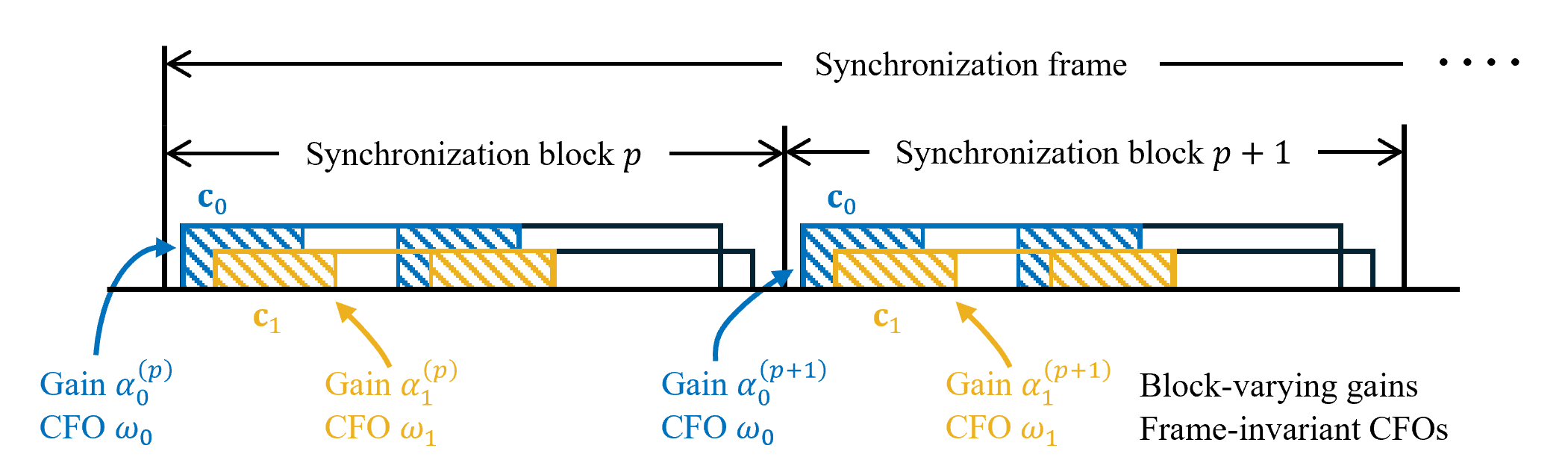}\caption{\label{fig:coherence_block_system}Hierarchical synchronization structure
under strong inter-cell interference.}
\end{figure}

\subsection{Signal Structure}

Consider a hierarchical synchronization structure with three levels:
a synchronization frame, synchronization blocks, and pilot sequences.\footnote{In 5G \ac{nr}, these correspond to the \ac{ss} burst set, \ac{ssb},
and \ac{pss}/\ac{sss} training sequences, respectively; see Appendix
\ref{sec:5GNR}.} A synchronization frame consists of $P$ synchronization blocks,
indexed by $p\in\mathcal{P}\triangleq\{0,1,\cdots,P-1\}$, and each
synchronization block contains two known pilot sequences. The overall
structure is illustrated in Fig.~\ref{fig:coherence_block_system}.

The received signal in each synchronization block is a superposition
of contributions from $K$ transmitters, indexed by $k\in\mathcal{K}\triangleq\{0,1,\cdots,K-1\}$.
For transmitter $k$, all synchronization blocks within one \ac{cfo}-coherent
interval share a common \ac{cfo}, as justified by the kinematic coherence
analysis in Section~\ref{sec:Kinematic_analysis}, while the complex
channel gain remains block-dependent.

\subsubsection{Pilot Signal Model}

Let $\mathcal{C}$ denote the set of all length-$N$ time-domain pilot
sequences. The synchronization signal transmitted by $k$ consists
of two pilot sequences separated by a fixed guard interval and is
written as
\begin{equation}
\mathbf{c}_{k}=\mathbf{c}_{k,0}\oplus\mathbf{0}\oplus\mu_{k}\mathbf{c}_{k,1}\label{eq:def_frame}
\end{equation}
where $\mathbf{c}_{k,0},\mathbf{c}_{k,1}\in\mathcal{C}$ denote the
selected pilot sequences, $\mathbf{0}$ is a length-$\tau_{0}$ all-zero
vector, and $\mu_{k}\in\mathbb{R}^{+}$ models the relative scaling
between the two pilot sequences. Individual pilot sequences may be
reused across transmitters, but the ordered pair ($\mathbf{c}_{k,0},\mathbf{c}_{k,1}$),
and thus the composite synchronization signal $\mathbf{c}_{k}$, uniquely
identifies transmitter $k$.

For two pilot sequences $\mathbf{s}_{k},\mathbf{s}_{l}\in\mathcal{C}$,
define the cross-correlation with timing offset $\tau$ and frequency
offset $\omega$ as 
\begin{equation}
r_{k,l}\left(\tau,\omega\right)\triangleq\frac{1}{N}\sum_{m=0}^{N-1}s_{k}\left[m\right]s_{l}^{*}\left[m-\tau\right]e^{-j\omega m}.\label{eq:def_ac}
\end{equation}
Since the pilot sequences are not perfectly orthogonal, we model the
cross-correlation for any mismatch ($k\neq l$) or nonzero delay ($\tau\neq0$)
as zero-mean circular complex Gaussian with a variance proportional
to $1/N$. For the matched case ($k=l$, $\tau=0$), the statistic
reduces to the sequence autocorrelation $r_{k}(\omega)$. Accordingly,
\begin{equation}
r_{k,l}\left(\tau,\omega\right)=\begin{cases}
r_{k}\left(\omega\right) & k=l,\tau=0\\
\mathcal{CN}\left(0,\sigma_{\text{c}}^{2}/N\right) & \text{otherwise}
\end{cases}\label{eq:dis_ac}
\end{equation}
where $\sigma_{\text{c}}^{2}$ captures sequence-family sidelobes
and implementation impairments ({\em e.g.}, filtering, residual multipath),
and the autocorrelation magnitude approximates $|r_{k}(\omega)|\approx\sin(N\omega/2)/(N\sin(\omega/2))|$,
\emph{e.g.}, for the m-sequence-based PSS and Gold-sequence-based
SSS in 5G \ac{nr}. Therefore, $|r_{k}(0)|=1$ and the envelope of
$|r_{k}(\omega)|$ decreases over $\left|\omega\right|$.

\subsubsection{Received Signal Model}

For transmitter $k$, let $\tau_{k}$ and $\omega_{k}$ denote the
timing and frequency offsets respectively, and let $\alpha_{k}^{(p)}$
denote the complex channel gain associated in synchronization block
$p$. Under the constant-amplitude, linear-phase channel model established
in Section~\ref{subsec:Channel_model}, the received signal at time
index $m$ is\footnote{Extension to frequency-selective fading is straightforward via multipath
delay and gain estimation. Here, we focus on the single-path (frequency-flat)
case, which is sufficient for the considered low-altitude scenarios.}
\begin{equation}
y\left[m\right]=\sum_{k=0}^{K-1}\sum_{p=0}^{P-1}\alpha_{k}^{\left(p\right)}c_{k}\left[m-pN_{\text{f}}-\tau_{k}\right]e^{-j\omega_{k}m}+\nu\left[m\right]\label{eq:def_rx_signal}
\end{equation}
for $m\in\{0,1,\cdots,N_{\text{o}}-1\}$, where $N_{\text{f}}$ is
the synchronization block length and $\nu[m]\sim\mathcal{CN}(0,\sigma_{\text{n}}^{2})$
is additive white Gaussian noise. The parameter $N_{\text{o}}$ is
the total observation length. Here, $c_{k}[m]$ is zero outside the
support of the corresponding synchronization signal. For notational
convenience, we collect the received samples into a vector 
\begin{equation}
\mathbf{y}\triangleq\left[y\left[0\right],y\left[1\right],\cdots,y\left[N_{\text{o}}-1\right]\right]^{\text{T}}\in\mathbb{C}^{N_{\text{o}}\times1}.\label{eq:def_y_vec}
\end{equation}

The timing offsets $[\tau_{k}]$ are estimated and the roots $\{\mathbf{c}_{k,0},\mathbf{c}_{k,1}\}$
are identified during the pre-processing stage, as detailed in Appendix~\ref{sec:5GNR}.
Accordingly, the following part focuses on estimating the \acpl{cfo}
$\boldsymbol{\omega}\triangleq[\omega_{k}]_{k\in\mathcal{K}}$, the
scaling factors $\boldsymbol{\mu}\triangleq[\mu_{k}]_{k\in\mathcal{K}}$,
and the block-dependent channel gains $\boldsymbol{\alpha}\triangleq[\alpha_{k}^{(p)}]_{k\in\mathcal{K},p\in\mathcal{P}}$.

\section{Kinematic \ac{cfo} Coherence Model\label{sec:Kinematic_analysis}}

The effective frequency offset $f$ in the channel model is locally
approximated as constant over a short interval. We now quantify the
kinematic conditions under which this approximation holds.

The frequency offset $f(t)$ of the dominant received component is
affected by both intrinsic oscillator mismatch and platform motion.
We write $f\left(t\right)=f_{\text{osc}}\left(t\right)+f_{d}\left(t\right)$,
where $f_{\text{osc}}\left(t\right)$ denotes the oscillator-mismatch-induced
offset and $f_{d}\left(t\right)$ denotes the motion-induced Doppler
shift. Over a short interval $[t_{0},t_{0}+\delta_{t}]$, $f_{\text{osc}}\left(t\right)$
is assumed to vary slowly enough to be approximated as constant, so
the \ac{cfo} variation is mainly determined by the Doppler term.
Accordingly, the \ac{cfo} drift over $[t_{0},t_{0}+\delta_{t}]$
is defined as 
\[
\Delta_{f}\triangleq\max_{t_{1},t_{2}\in\left[t_{0},t_{0}+\delta_{t}\right]}\left|f_{d}\left(t_{1}\right)-f_{d}\left(t_{2}\right)\right|.
\]
An interval satisfying $\Delta_{f}\le\bar{\delta}_{f}$ is referred
to as a {\em \ac{cfo}-coherence frame}, where $\bar{\delta}_{f}$
is a prescribed drift tolerance.

Let $\mathbf{v}\left(t\right)$ and $\mathbf{a}\left(t\right)$ denote
the instantaneous velocity and acceleration of the \ac{uav}, respectively,
$d\left(t\right)$ the effective path length of the dominant component,
and $\theta\left(t\right)$ the angle between $\mathbf{v}\left(t\right)$
and its arrival direction. Suppose that, for all $t\in\left[t_{0},t_{0}+\delta_{t}\right]$
\[
\left\Vert \mathbf{v}\left(t\right)\right\Vert \le\bar{v},\left\Vert \mathbf{a}\left(t\right)\right\Vert \le\bar{a},\sin\theta\left(t\right)\le\bar{s},d\left(t\right)\ge\underline{d}.
\]
Given a drift tolerance $\bar{\delta}_{f}$, the following condition
ensures \ac{cfo} coherence.
\begin{theorem}[Kinematic condition for \ac{cfo}
coherence]
\label{thm:Kinematic_condition}
The interval $[t_{0},t_{0}+\delta_{t}]$ forms a \ac{cfo}-coherence
frame, i.e., $\Delta_{f}\le\bar{\delta}_{f}$, if 
\begin{equation}
\bar{a}+\frac{\bar{v}^{2}\bar{s}}{\underline{d}}\le\frac{\lambda_{\text{c}}\bar{\delta}_{f}}{\delta_{t}}.\label{eq:Kinematic_condition}
\end{equation}
\end{theorem}
\begin{IEEEproof}
See Appendix \ref{sec:proof_lem_cfo_drift}.
\end{IEEEproof}
As an example, consider a 5G system with carrier frequency $f_{\text{c}}=3\text{ GHz}$
and one \ac{ss} burst observation duration $\delta_{t}=2.5\text{ ms}$.
For a drift tolerance $\bar{\delta}_{f}=1\text{ Hz}$, the resulting
kinematic budget is $40\text{ }\mathrm{m/s^{2}}$. Under steady flight
($\bar{a}\approx0$) at distance $\underline{d}=50\text{ m}$, the
\ac{cfo}-coherence constraint allows tangential velocities up to
$44\text{ m/s}$. Even under wind-induced stabilization maneuvers
(\emph{e.g.}, $\bar{a}=5\mathrm{\text{ }m/s^{2}}$), the remaining
margin permits velocities up to $41\mathrm{\text{ }m/s}$, which is
still well above typical \ac{uav} operating speeds. These results
indicate that the proposed \ac{cfo}-coherence condition is readily
satisfied in practical \ac{uav} deployments, even under moderate
flight perturbations.

\section{Multi-Block Processing for Joint CFO-Channel Estimation\label{sec:Estimation}}

This section develops a multi-block framework for joint \ac{cfo}
and channel estimation. We first develop the channel estimator, then
the \ac{cfo} estimator, and finally the joint iterative algorithm.

\subsection{Multi-Block Processing for Channel Estimation}

Denote the received signal in synchronization block $p\in\mathcal{P}$
as $\mathbf{y}^{(p)}\triangleq[y\left[m+pN_{\text{f}}\right]]_{m\in\{0,1,\cdots,N_{\text{f}}-1\}}\in\mathbb{C}^{N_{\text{f}}\times1}$.
Under the signal model in (\ref{eq:def_rx_signal}), $\mathbf{y}^{(p)}$
can be written as 
\[
\mathbf{y}^{(p)}=\underbrace{\left(\boldsymbol{\Omega}\left(\boldsymbol{\omega}\right)\varodot\mathbf{C}\left(\boldsymbol{\mu}\right)\right)\mathbf{D}^{(p)}\left(\boldsymbol{\omega}\right)}_{\triangleq\boldsymbol{A}^{(p)}\left(\boldsymbol{\omega},\boldsymbol{\mu}\right)\in\mathbb{C}^{N_{\text{f}}\times K}}\boldsymbol{\alpha}^{\left(p\right)}+\mathbf{v}^{(p)}
\]
where $\varodot$ denotes the Hadamard product of compatible matrices,
$\mathbf{C}\left(\boldsymbol{\mu}\right)=[\mathbf{0}_{\tau_{k}\times1};\mathbf{c}_{k};\mathbf{0}]_{k\in\mathcal{K}}\in\mathbb{C}^{N_{\text{f}}\times K}$,
$\boldsymbol{\Omega}\left(\boldsymbol{\omega}\right)=[\Omega_{k}(\omega_{k})]_{k\in\mathcal{K}}\in\mathbb{C}^{N_{\text{f}}\times K}$,
$\Omega_{k}(\omega_{k})=[1;e^{-j\omega_{k}};\cdots;e^{-j\omega_{k}(N_{\text{f}}-1)}]\in\mathbb{C}^{N_{\text{f}}\times1}$,
$\mathbf{D}^{(p)}\left(\boldsymbol{\omega}\right)=\text{diag}([e^{-j\omega_{k}pN_{\text{f}}}]_{k\in\mathcal{K}})\in\mathbb{C}^{K\times K}$,
$\boldsymbol{\alpha}^{(p)}=[\alpha_{k}^{(p)}]_{k\in\mathcal{K}}\in\mathbb{C}^{K\times1}$,
and $\mathbf{v}^{(p)}=[v[m+pN_{\text{f}}]]_{m\in\{0,\cdots,N_{\text{f}}-1\}}\in\mathbb{C}^{N_{\text{f}}\times1}$.

Stacking the received signals from all synchronization blocks yields
\begin{equation}
\mathbf{y}=\underbrace{\text{blkdiag}\left(\boldsymbol{A}^{(0)},\cdots,\boldsymbol{A}^{(P-1)}\right)}_{\triangleq\boldsymbol{A}\left(\boldsymbol{\omega},\boldsymbol{\mu}\right)\in\mathbb{C}^{N_{\text{f}}P\times KP}}\boldsymbol{\alpha}+\boldsymbol{v}.\label{eq:y_matrix_form_1}
\end{equation}
where $\text{blkdiag}(\cdot)$ denotes the block-diagonal operator.

Accordingly, given $\left(\boldsymbol{\omega},\boldsymbol{\mu}\right)$,
the channel gains are obtained from the least-squares problem as
\begin{equation}
\hat{\boldsymbol{\alpha}}=\boldsymbol{\boldsymbol{A}\left(\boldsymbol{\omega},\boldsymbol{\mu}\right)}^{\dagger}\mathbf{y}\label{eq:est_alpha}
\end{equation}
where $(\cdot)^{\dagger}$ denotes the Moore-Penrose pseudoinverse.

Using the two-pilot structure, the received signal $\mathbf{y}$ can
be rewritten as 
\[
\mathbf{y}=\boldsymbol{A}_{0}\left(\boldsymbol{\omega}\right)\boldsymbol{\alpha}+\underbrace{\left[\begin{array}{c}
\boldsymbol{A}_{1}^{(0)}\text{diag}\left(\boldsymbol{\alpha}^{\left(0\right)}\right)\\
\vdots\\
\boldsymbol{A}_{1}^{(P-1)}\text{diag}\left(\boldsymbol{\alpha}^{\left(P-1\right)}\right)
\end{array}\right]}_{\triangleq\boldsymbol{B}\left(\boldsymbol{\omega},\boldsymbol{\alpha}\right)\in\mathbb{C}^{N_{\text{f}}P\times K}}\boldsymbol{\mu}+\boldsymbol{v}
\]
where $\boldsymbol{A}_{i}=\text{blkdiag}(\boldsymbol{A}_{i}^{(0)},\cdots,\boldsymbol{A}_{i}^{(P-1)})\in\mathbb{C}^{N_{\text{f}}P\times KP}$,
$\boldsymbol{A}_{i}^{(p)}=(\boldsymbol{\Omega}(\boldsymbol{\omega})\varodot\mathbf{C}_{i})\mathbf{D}^{(p)}\left(\boldsymbol{\omega}\right)\in\mathbb{C}^{N_{\text{f}}\times K}$,
$\mathbf{C}_{i}=[\mathbf{0}_{(\tau_{k}+\tau_{\text{c}}i)\times1};\mathbf{c}_{k,i};\mathbf{0}]_{k\in\mathcal{K}}\in\mathbb{C}^{N_{\text{f}}\times K}$,
and $\tau_{\text{c}}=N+\tau_{0}$.

Given $\left(\boldsymbol{\omega},\boldsymbol{\alpha}\right)$, the
scaling factors are then obtained by least squares as
\begin{equation}
\hat{\boldsymbol{\mu}}=\boldsymbol{B}\left(\boldsymbol{\omega},\boldsymbol{\alpha}\right)^{\dagger}\left(\mathbf{y}-\boldsymbol{A}_{0}\left(\boldsymbol{\omega}\right)\boldsymbol{\alpha}\right).\label{eq:est_mu}
\end{equation}

\subsection{Multi-Block Processing for CFO Estimation}

We first characterize the correlation between the received signal
and the reference signals. For synchronization block $p$, define
the normalized cross-correlation between the received samples and
the $i$th synchronization sequences of transmitter $k$, \emph{i.e.},
$\mathbf{c}_{k,0}$ and $\mathbf{c}_{k,1}$, as 
\begin{equation}
r_{y,k}^{p,i}\triangleq\frac{1}{N}\sum_{m=\tau_{k}}^{\tau_{k}+N-1}y\left[m+pN_{\text{f}}+i\tau_{\text{c}}\right]c_{k,i}^{\text{H}}\left[m-\tau_{k}\right]\label{eq:def_x_r_yk}
\end{equation}
where $k\in\mathcal{K}$, $p\in\mathcal{P}$, and $i\in\{0,1\}$.

Then, the following lemma characterizes its distribution.
\begin{lemma}[Cross-correlation distribution]
\label{lem:x_r_yk}The normalized cross-correlation $r_{y,k}^{p,i}$
satisfies 
\begin{equation}
r_{y,k}^{p,i}\sim\alpha_{k}^{(p)}r_{k}\left(\omega_{k}\right)\text{\ensuremath{\left(\mu_{k}e^{-j\omega_{k}\tau_{\text{c}}}\right)}}^{i}+\mathcal{CN}\left(0,\sigma_{k,p,i}^{2}\right)\label{eq:rst_x_r_yk}
\end{equation}
where the variance $\sigma_{k,p,i}^{2}$ collects the averaged interference-plus-noise
power, \emph{i.e.}, 
\[
\sigma_{k,p,i}^{2}=\frac{\sum_{q\neq k,q\in\mathcal{K}}\left(\mu_{q}^{i}\alpha_{q}^{\left(p\right)}\sigma_{c}\right)^{2}+\sigma_{n}^{2}}{N}.
\]
\end{lemma}
\begin{IEEEproof}
See Appendix \ref{sec:proof_lem_x_r_yk}.
\end{IEEEproof}
Lemma~\ref{lem:x_r_yk} shows that $r_{y,k}^{p,i}$ is a noisy observation
of the common term $\alpha_{k}^{(p)}r_{k}\left(\omega_{k}\right)$,
scaled by $(\mu_{k}e^{-j\omega_{k}\tau_{\text{c}}})^{i}$ associated
with the \ac{cfo} $\omega_{k}$. In particular, for fixed $(k,p)$,
the pair $(r_{y,k}^{p,0},r_{y,k}^{p,1})$ shares the same unknown
complex gain $\alpha_{k}^{(p)}r_{k}\left(\omega_{k}\right)$ while
the \ac{cfo} only appears through a deterministic phase rotation
between the two correlations. Consequently, taking the ratio between
$r_{y,k}^{p,0}$ and $r_{y,k}^{p,1}$ eliminates the common gain and
isolates the \ac{cfo}-dependent phase term.

More precisely, taking expectations in (\ref{eq:rst_x_r_yk}) yields
$\mathbb{E}[r_{y,k}^{p,1}]/(\mathbb{E}[r_{y,k}^{p,0}])=\mu_{k}e^{-j\omega_{k}\tau_{\text{c}}}$
and the phase of this ratio uniquely determines the \ac{cfo}. This
observation leads to the following single-block \ac{cfo} estimator
\begin{equation}
\hat{\omega}_{k}^{\text{single}}=\frac{1}{\tau_{\text{c}}}\angle\left(\mathbb{E}\left[r_{y,k}^{p,1}\right]\left(\mathbb{E}\left[r_{y,k}^{p,0}\right]\right)^{\text{H}}\right)\label{eq:single_est_omega}
\end{equation}
since $\mu_{k}$ is real.

To exploit all synchronization blocks within the \ac{cfo}-coherent
frame jointly, we collect the correlation outputs associated with
\ac{bs} $k$ into the vector $\mathbf{r}_{y,k}=[r_{y,k}^{p,i}]_{p\in\mathcal{P},i\in\{0,1\}}$,
and define the multi-block likelihood as $\mathbb{P}(\mathbf{r}_{y,k}|\omega_{k})$.
Since synchronization blocks are non-overlapping in time as analyzed
in Appendix~\ref{sec:5GNR}, the correlation outputs are conditionally
independent given $\omega_{k}$, yielding 
\[
\mathbb{P}\left(\mathbf{r}_{y,k}|\omega_{k}\right)=\prod_{p\in\mathcal{P}}\prod_{i\in\{0,1\}}\mathbb{P}\left(r_{y,k}^{p,i}|\omega_{k}\right).
\]

According to Lemma~\ref{lem:x_r_yk}, each $r_{y,k}^{p,i}$ is complex
Gaussian, so the log-likelihood function reduces (up to an additive
constant) becomes
\begin{align}
 & \ln\left(\mathbb{P}\left(\mathbf{r}_{y,k}|\omega_{k}\right)\right)\nonumber \\
 & \propto-\sum_{p\in\mathcal{P}}\sum_{i\in\left\{ 0,1\right\} }\frac{\left|r_{y,k}^{p,i}-\alpha_{k}^{(p)}r_{k}\left(\omega_{k}\right)\text{\ensuremath{\left(\mu_{k}e^{-j\omega_{k}\tau_{\text{c}}}\right)}}^{i}\right|^{2}}{\sigma_{k,p,i}^{2}}.\label{eq:def_f}
\end{align}

Maximizing this log-likelihood with respect to $\omega_{k}$ leads
to the following closed-form multi-block \ac{ml} \ac{cfo} estimator.
\begin{proposition}[\Ac{cfo} estimator]
\label{prop:joint_est_omega}The relaxed multi-block \ac{ml} \ac{cfo}
estimator is 
\begin{equation}
\hat{\omega}_{k}^{\text{multi}}=\frac{1}{\tau_{c}}\angle\left\{ \psi_{1}\psi_{0}^{\text{H}}\right\} \label{eq:joint_est_omega}
\end{equation}
where 
\begin{equation}
\psi_{0}\triangleq\sum_{p\in\mathcal{P}}\frac{\alpha_{k}^{(p)}\left(r_{y,k}^{p,0}\right)^{\text{H}}}{\sigma_{k,p,0}^{2}},\quad\psi_{1}\triangleq\sum_{p\in\mathcal{P}}\frac{\alpha_{k}^{(p)}\mu_{k}\left(r_{y,k}^{p,1}\right)^{\text{H}}}{\sigma_{k,p,1}^{2}}.\label{eq:def_phi1}
\end{equation}
This estimator is exact for the original \ac{ml} problem when
\begin{equation}
r_{k}\left(\omega_{k}\right)=\frac{|\psi_{0}|+\left|\psi_{1}\right|}{\Gamma_{0}+\Gamma_{1}}e^{-j\angle\psi_{0}}\triangleq\gamma_{k}\label{eq:opt_c_est}
\end{equation}
where $\Gamma_{0}$and $\Gamma_{1}$ are aggregated average \ac{sinr}
defined as
\begin{equation}
\Gamma_{0}=\sum_{p\in\mathcal{P}}\frac{\left|\alpha_{k}^{(p)}\right|^{2}}{\sigma_{k,p,0}^{2}},\,\Gamma_{1}=\sum_{p\in\mathcal{P}}\frac{\left|\mu_{k}\alpha_{k}^{(p)}\right|^{2}}{\sigma_{k,p,1}^{2}}.\label{eq:def_gamma_c}
\end{equation}
\end{proposition}
\begin{IEEEproof}
See Appendix \ref{sec:proof_prop_joint_est}.
\end{IEEEproof}
Proposition~\ref{prop:joint_est_omega} shows that the estimator
coherently combines all synchronization blocks within the \ac{cfo}-coherence
frame through weighted sum of the correlation outputs (via $\psi_{0}$
and $\psi_{1}$), where the weight of each block $p$ is proportional
to its effective average \ac{sinr}, $\alpha_{k}^{(p)}/\sigma_{k,p,i}^{2}$.
Thus, blocks with higher reliability contribute more strongly to the
final \ac{cfo} estimate. In this sense, the estimator performs coherent
phase aggregation across synchronization blocks, effectively increasing
the observation length available for \ac{cfo} estimation.

\begin{figure}
\centering{}\includegraphics[width=1\columnwidth]{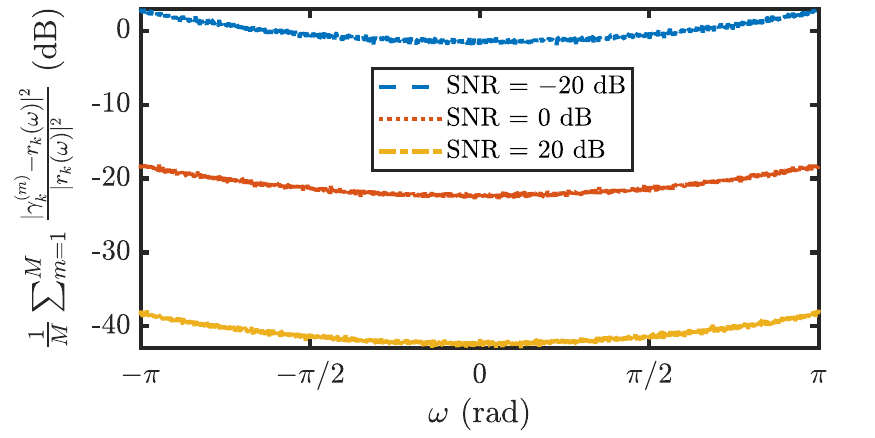}\caption{\label{fig:mismatch_r_gamma}Verification of the relaxed-\ac{ml}
equivalence condition. The feasibility mismatch is approximately on
the order of $1/(N\cdot\text{SNR})$.}
\end{figure}

Fig.~\ref{fig:mismatch_r_gamma} verifies the feasibility condition
of the relaxed estimator using the \ac{nmse}. A small \ac{nmse}
indicates that the relaxed optimum approximately satisfies the constrained
\ac{ml} condition in (\ref{eq:opt_c_est}). The results show that
the \ac{nmse} scales approximately as $1/(N\cdot\text{SNR})$, which
confirms that the relaxed estimator closely approximates the optimal
\ac{ml} estimator at moderate \ac{snr} levels, \emph{e.g.}, around
$-20\text{ dB}$ \ac{nmse} at $0\text{ dB}$ \ac{snr}.

\subsection{Joint CFO-Channel Estimation Algorithm}

\begin{algorithm}[t]
\caption{\label{alg:estimation} Joint \ac{cfo} and Channel Estimation Algorithm}

\textbf{\# Initialization}: $\hat{\boldsymbol{\omega}}\leftarrow\mathbf{0}$,
$\hat{\boldsymbol{\alpha}}\leftarrow\mathbf{0}$, $\hat{\boldsymbol{\mu}}\leftarrow\mathbf{1}$,
$\tilde{\mathbf{y}}\leftarrow\mathbf{y}$
\begin{enumerate}
\item \label{alg_step:update_channel}Channel Estimation: Construct $\mathbf{A}$,
$\mathbf{A}_{0}$, and $\mathbf{B}$, then update channel parameters
$\hat{\boldsymbol{\alpha}}$ and $\hat{\boldsymbol{\mu}}$ based on
multi-block estimator (\ref{eq:est_alpha}) and (\ref{eq:est_mu}).
\item SIC-based \ac{cfo} Estimation: For each \ac{bs} $k$, perform
\ac{sic} and compensation: $\tilde{\mathbf{y}}_{k}\leftarrow(\mathbf{y}-\boldsymbol{A}(\hat{\boldsymbol{\tau}}_{-k},\hat{\boldsymbol{\omega}}_{-k})\hat{\boldsymbol{\alpha}}_{-k})\varodot\Omega_{k}(\omega_{k})$.
Then, estimate the remaining \ac{cfo} $\delta_{k}$ based on the
multi-block estimator (\ref{eq:joint_est_omega}), and update $\hat{\omega}_{k}\leftarrow\hat{\omega}_{k}+\delta_{k}$.
\item Go to step \ref{alg_step:update_channel}) until $\sum_{k}\left|\delta_{k}\right|<\epsilon$.
\end{enumerate}
\end{algorithm}

The proposed algorithm alternates between channel-parameter update
and \ac{cfo} refinement. Given the current estimates $\left(\hat{\boldsymbol{\tau}},\hat{\boldsymbol{\omega}}\right)$,
the block-dependent channel gains $\hat{\boldsymbol{\alpha}}$ and
transmitter-dependent scaling factors $\hat{\boldsymbol{\mu}}$ are
first updated via (\ref{eq:est_alpha}) and (\ref{eq:est_mu}). Then,
for each transmitter $k$, we form the \ac{sic}-processed and \ac{cfo}-precompensated
waveform 
\begin{equation}
\tilde{\mathbf{y}}_{k}=\Bigl(\mathbf{y}-\boldsymbol{A}\bigl(\hat{\boldsymbol{\tau}}_{-k},\hat{\boldsymbol{\omega}}_{-k}\bigr)\hat{\boldsymbol{\alpha}}_{-k}\Bigr)\odot\boldsymbol{\Omega}_{k}(\hat{\omega}_{k})\label{eq:residual-signal}
\end{equation}
where ``$-k$'' denotes the collection of parameters for all transmitters
except $k$. Based on $\tilde{\mathbf{y}}_{k}$, a residual \ac{cfo}
increment $\delta_{k}$ is estimated using the multi-block estimator
in (\ref{eq:joint_est_omega}), and the \ac{cfo} is updated as $\hat{\omega}_{k}\leftarrow\hat{\omega}_{k}+\delta_{k}$.
The iterations terminate when the residual \ac{cfo} updates become
negligible. The single-block case is recovered by setting $P=1$.

\section{Performance Analysis\label{sec:Performance-Guarantee}}

This section analyzes the estimation accuracy of the proposed multi-block
algorithm and clarifies the effects of synchronization structure and
platform kinematics on the achievable performance.
\begin{proposition}[Asymptotic error bound]
\label{prop:crlb} The \ac{cfo} estimator $\hat{\omega}_{k}^{\text{multi}}$ is asymptotically distributed
as
\begin{equation}
\hat{\omega}_{k}^{\text{multi}}\stackrel{a}{\sim}\mathcal{N}\left(\omega_{k},V_{k}\right)\label{eq:est_omega_multi}
\end{equation}
where $V_{k}$ is the \ac{crlb} given by
\begin{equation}
V_{k}=\frac{\Gamma_{0}^{-1}+\Gamma_{1}^{-1}}{2\tau_{\text{c}}^{2}\left|\gamma_{k}\right|^{2}}.\label{eq:crlb_c}
\end{equation}
The approximation remainder is of stochastic order $O_{p}((\Gamma_{0}^{-1}+\Gamma_{1}^{-1})/(\tau_{\text{c}}|\gamma_{k}|^{2})).$
\end{proposition}
\begin{IEEEproof}
See Appendix \ref{sec:proof_prop_crlb}.
\end{IEEEproof}
Proposition~\ref{prop:crlb} shows that the \ac{crlb} degrades with
increasing interference (through $\sigma_{k,p,i}^{2}$) and with decreasing
$|\gamma_{k}|^{2}$ (typically when $|\omega_{k}|$ is large). Therefore,
a coarse-to-fine refinement strategy in step 2 of Algorithm~\ref{alg:estimation},
consisting of initial estimation, pre-compensation, successive interference
cancellation, and re-estimation, can progressively improve the effective
\ac{sinr} and reduce the residual offset, thereby lowers the achievable
error bound.

Moreover, the variance of the \ac{cfo} estimator (\ref{eq:est_omega_multi})
is lower-bounded by a term inversely proportional to the aggregated
\ac{sinr} over all synchronization blocks. The quantities $\Gamma_{0}$
and $\Gamma_{1}$ collect, for each synchronization sequence, the
per-block received average \ac{sinr}, $|\alpha_{k}^{(p)}|^{2}/\sigma_{k,p,i}^{2}$.
As a result, blocks affected by strong interference contribute little
to the aggregated information, whereas high-\ac{sinr} blocks dominate
the estimation accuracy. This implies that accurate \ac{cfo} estimation
for a weak block is still possible as long as at least one block within
the same \ac{cfo}-coherence interval has sufficiently high \ac{sinr}.
This dependence becomes explicit under the equal-\ac{sinr} scenario
considered next.
\begin{theorem}[Scaling law]
\label{the:crlb_over_P}If all synchronization blocks experience the
same \ac{sinr}, that is $|\alpha_{k}^{(p)}|^{2}/\sigma_{k,p,i}^{2}\triangleq\rho_{i},\forall p\in\mathcal{P}$, then the \ac{crlb} for $\omega_{k}$ becomes
\begin{equation}
V_{k}=\frac{\rho_{0}^{-1}+\rho_{1}^{-1}}{2P\tau_{\text{c}}^{2}\left|\gamma_{k}\right|^{2}}\propto\frac{1}{P}.\label{eq:crlb_c_P}
\end{equation}
\end{theorem}
\begin{IEEEproof}
The aggregated becomes \ac{sinr} $\Gamma_{i}=P\rho_{i}$, then substituting it into (\ref{eq:crlb_c}) gives
the result.
\end{IEEEproof}
Hence, the minimum achievable variance decreases proportionally to
$1/P$ as the number of synchronization blocks $P$ increases. This
result highlights the key advantage of multi-block processing: coherent
aggregation across synchronization blocks effectively increases the
observation length and improves \ac{cfo} estimation accuracy. As
more spatial beams are probed within each synchronization frame, the
proposed estimator can exploit a larger effective $P$ and achieve
proportionally lower estimation variance.

Multiple synchronization frames can also be jointly processed when
they fall within the same \ac{cfo}-coherent interval, further increasing
the number of coherent blocks. The next result relates the effective
block count to the platform kinematics.
\begin{theorem}[Kinematic scaling law]
\label{the:cogerance_sampling_setting}Denote $\bar{\delta}_{\text{f}}$
as the allowable \ac{cfo} coherence limit, the effective number of
synchronization blocks can be coherently exploited is 
\[
P_{\text{eff}}\approx\frac{\bar{\delta}_{\text{f}}\lambda_{\text{c}}P}{\left(\bar{a}+\frac{\bar{v}^{2}\bar{s}}{\underline{d}}\right)\tau_{\text{p}}}
\]
where $\tau_{\text{p}}$ is the synchronization-frame period.

Under the same-\ac{sinr} assumption of Theorem~\ref{the:crlb_over_P},
the resulting drift-limited \ac{crlb} for $\omega_{k}$ satisfies
\[
V_{k}\approx\frac{\tau_{\text{p}}\left(\rho_{0}^{-1}+\rho_{1}^{-1}\right)}{2P\lambda_{\text{c}}\bar{\delta}_{\text{f}}\tau_{\text{c}}^{2}\left|r_{k}\left(\omega_{k}\right)\right|^{2}}\left(\bar{a}+\frac{\bar{v}^{2}\bar{s}}{\underline{d}}\right).
\]
\end{theorem}
\begin{IEEEproof}
From Theorem~1, the \ac{cfo} coherence time is $\delta_{t}\approx\lambda_{\text{c}}\bar{\delta}_{f}/(\bar{a}+\bar{v}^{2}\bar{s}/\underline{d})$.
Since each synchronization frame contributes $P$ synchronization
blocks, the effective number of coherent blocks is $P_{\text{eff}}\approx\delta_{t}P$.
Substituting $P_{\text{eff}}$ into Theorem~\ref{the:crlb_over_P}
gives the result.
\end{IEEEproof}
Theorem~\ref{the:cogerance_sampling_setting} shows that the drift-limited
\ac{crlb} scales linearly with the kinematic budget $\bar{a}+\bar{v}^{2}\bar{s}/\underline{d}$.
The first term shows that rapid acceleration or maneuvering directly
degrades \ac{cfo} accuracy. The second term captures the geometry-dependent
motion penalty: it grows quadratically with speed $\bar{v}$, increases
with the maximum tangentiality factor $\bar{s}$, and decreases with
the minimum propagation distance $\underline{d}$. Therefore, \ac{cfo}-sensitive
measurements should favor smooth trajectories, moderate speed, and
near-radial motion, particularly when the platform operates close
to the transmitter. In the idealized zero-drift limit, the effective
coherent interval becomes unbounded and the \ac{crlb} tends to zero.

Similar scaling also applies to the approximation remainder in Proposition~\ref{prop:crlb}.
Hence, a larger number of coherent blocks and a smaller kinematic budget reduce the estimation variance and improve the validity of
the estimator-level Gaussian approximation.
\begin{figure}
\centering{}\includegraphics[width=0.9\columnwidth]{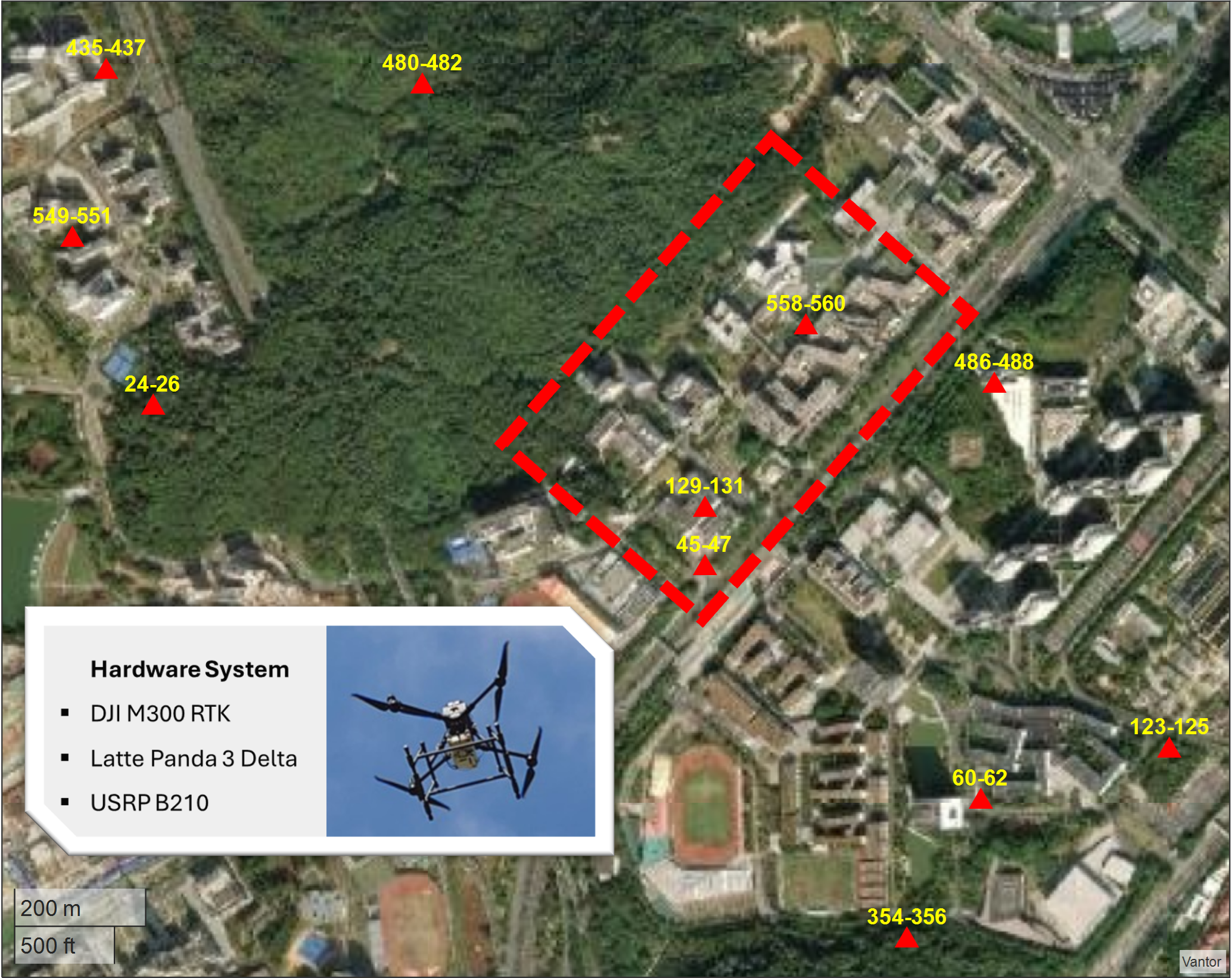}\caption{\label{fig:hardware_system}Aerial sampling over the CUHK-Shenzhen
campus. A DJI M300 RTK \ac{uav} equipped with an on-board Latte Panda~3
Delta computer, and a USRP~B210 (calibrated) software-defined radio
flies at an altitude of $150\text{ m}$ to collect measurements within
the area outlined by the dashed red polygon. Red triangles with yellow
labels indicate the locations and \acpl{pci} of a subset of known
terrestrial \acpl{bs}.}
\end{figure}
\begin{figure*}
\centering{}\includegraphics[width=1\textwidth]{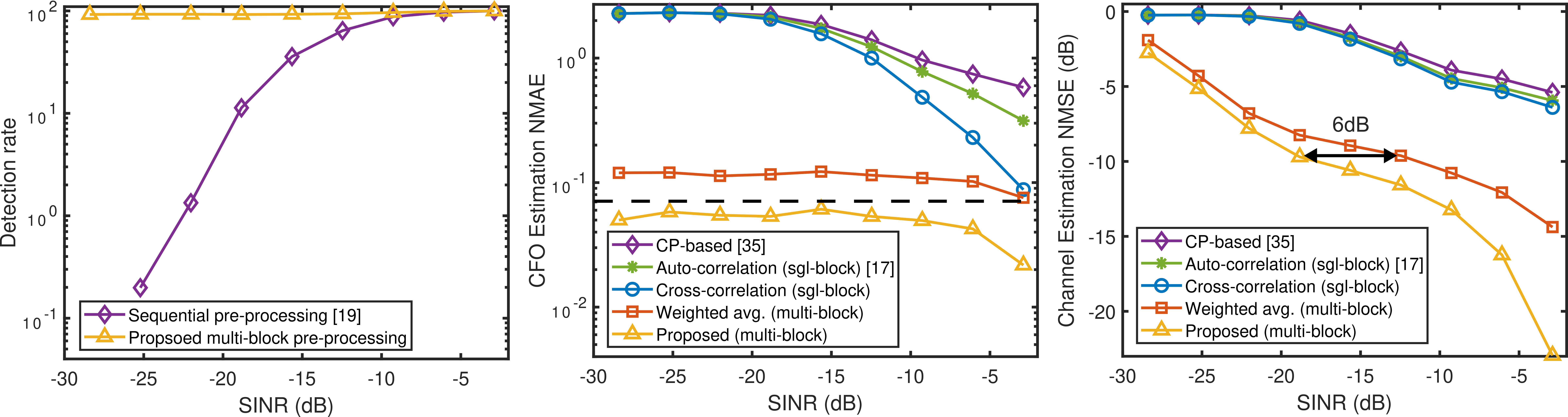}\caption{\label{fig:simulation_results_all}The detection rate, CFO estimation
MAE, and Channel estimation NMSE over SINR.}
\end{figure*}

\section{Performance Evaluation\label{sec:evaluation}}

In this section, both simulation and experimental platforms are developed to validate the proposed multi-block estimation algorithm. The simulation platform generates 5G \ac{nr} waveforms based on a configurable system model~\cite{3gpp:ts38211,3gpp:ts38213,3gpp:ts38331} and is used to evaluate the robustness of the algorithm under various controlled
scenarios. In contrast, the experimental platform collects real-world 5G waveforms in field deployments, enabling assessment of the algorithm's effectiveness in practical systems for radio map construction.

\subsection{Platform Settings}

\begin{figure}
\begin{centering}
\includegraphics[width=0.9\columnwidth]{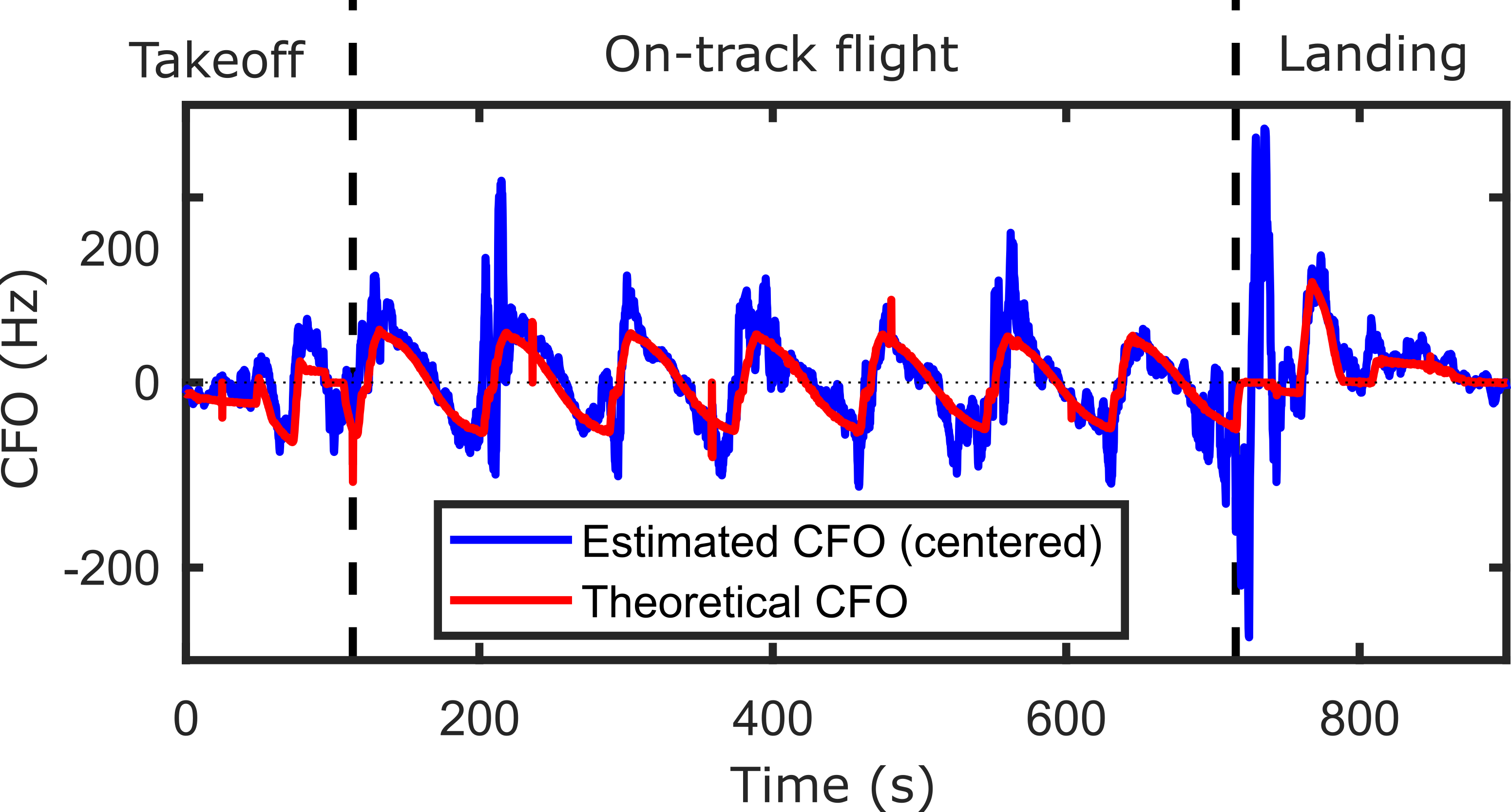}
\par\end{centering}
\caption{\label{fig:cfo_field_test}\ac{cfo} fluctuation over time during the low-altitude radio-map measurements, illustrating the correlation between the estimated \ac{cfo} (centered) and the theoretical \ac{cfo} across the three flight phases: takeoff, on-track flight, and landing.}
\end{figure}

\subsubsection{Simulation platform}

The simulation platform generates standard-compliant 5G \ac{nr} downlink waveforms. The frequency-domain sequences for the \ac{pss}, and \ac{sss} are generated according to the 3GPP specifications detailed in Appendix~\ref{sec:5GNR}, and the composite baseband signal is synthesized via \ac{ofdm} modulation. To emulate a low-altitude interference-limited scenario, we consider a configurable number of interfering \acpl{bs}. For each \ac{bs} $k$, the propagation delay $\tau_{k}\sim\mathcal{U}[0,\tau_{\text{max}}]$ and the \ac{cfo} $\omega_{k}\sim\mathcal{U}[-\omega_{\text{max}},\omega_{\text{max}}]$ model asynchronous network deployment and oscillator offsets, while the per-\ac{ssb} complex gain $\alpha_{k}^{(p)}\sim\mathcal{CN}(0,1)$ follows a Rayleigh fading model. The received signal is the superposition of all \acpl{ssb} from all \acpl{bs} plus complex white Gaussian noise $v\sim\mathcal{CN}(0,\delta_{n}^{2})$. Unless stated otherwise, each performance point is averaged over 200 Monte Carlo trials with $K=12$ \acpl{bs} and $P=12$ \acpl{ssb}.
\begin{figure*}
\centering{}\includegraphics[width=1\textwidth]{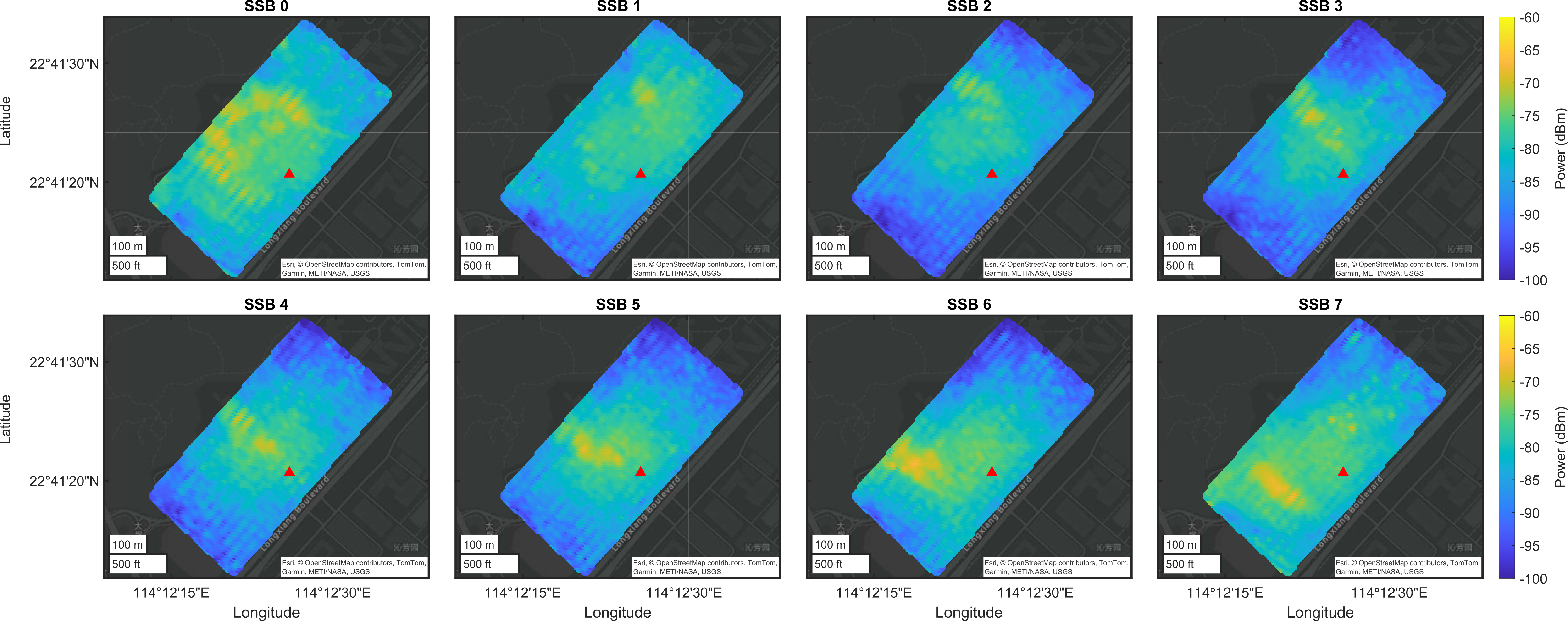}\caption{\label{fig:per_beam_radio_map}Per-beam low-altitude radio maps for
all beams 0-7 of the cell with \ac{pci} 45, where the red rectangle
marks the base-station location.}
\end{figure*}
\begin{figure*}
\centering{}\includegraphics[width=1\textwidth]{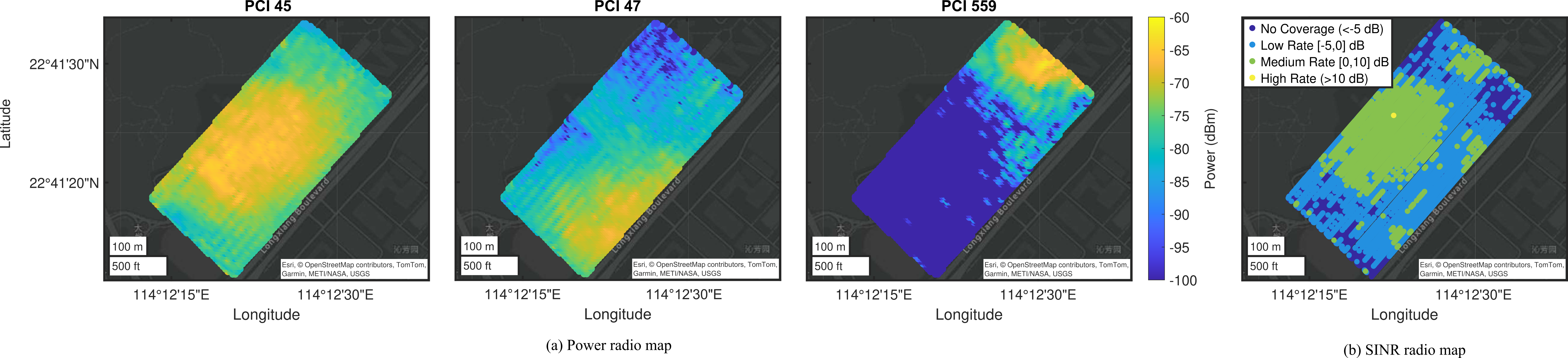}\caption{\label{fig:per_cell_radio_map}Per-cell low-altitude radio maps. (a)
Power radio maps for representative cells. (b) Integrated \ac{sinr}
radio map combining all detectable cells.}
\end{figure*}

\subsubsection{Experimental Platform}

A DJI M300 RTK \ac{uav} carries a \ac{usrp} to collect 5G \ac{nr}
signals, as illustrated in Fig.~\ref{fig:hardware_system}, with
carrier frequency $f_{c}=2.52495\text{ GHz}$ and bandwidth $20\text{ MHz}$.

\subsection{Simulation Results}

The detection performance of the proposed pre-processing algorithm is compared with the conventional sequential processing baseline (as
detailed in Appendix~\ref{sec:5GNR}) in Fig.~\ref{fig:simulation_results_all}(a). For both methods, the detection rate increases with the \ac{sinr}, but the proposed algorithm consistently outperforms the baseline over the entire \ac{sinr} range, with a significant gain in the low-\ac{sinr} regime. This robustness gain is attributed to the exploitation of the cross-frame structure, where we use the strong blocks as anchors to detect the weak blocks, thereby leading to robust detection.

The estimation performance of the proposed framework is compared with four baselines: i) a \ac{cp}-based blind estimator~\cite{SinKumMajSat:J25}; ii) an auto-correlation estimator that splits each synchronization sequence into two segments and obtains the \ac{cfo} from their phase difference~\cite{TunRivGarMel:J23}; iii) the single-block cross-correlation estimator in (\ref{eq:single_est_omega}); and iv) a multi-block baseline that forms a power-weighted average of the separately estimated \acpl{cfo}.

Fig.~\ref{fig:simulation_results_all}(b) shows the \ac{mae} of the \ac{cfo} (normalized by subcarrier spacing $30\text{ KHz}$)
estimate versus \ac{sinr}. Our method achieves a normalized \ac{cfo} error of approximately $0.05$, even at $-30\text{ dB}$ \ac{sinr}.
In \ac{ofdm} systems, keeping the residual \ac{cfo} well below $0.1$ subcarrier spacing is critical to minimize inter-carrier interference. This result confirms that our estimator provides sufficient accuracy for reliable symbol demodulation. Even in the single-block case, the proposed cross-correlation estimator outperforms auto-correlation because it exploits a longer phase baseline.

The improved \ac{cfo} estimation directly benefits subsequent channel estimation. As illustrated in Fig.~\ref{fig:simulation_results_all}(c), the \ac{nmse} of the channel estimate exhibits the same ordering as the \ac{cfo} \ac{mae}. This confirms that the more accurate \ac{cfo} compensation provided by the proposed framework is crucial for high-fidelity channel characterization.

\subsection{Field Test Results}

We first validate the proposed estimator and the overall measurement system by comparing the Doppler-induced theoretical \ac{cfo} drift
with the demeaned \ac{cfo} produced by our algorithm. According to the Doppler relation, the motion-induced \ac{cfo} offset satisfies
$\Delta f\approx-v_{\text{LOS}}f_{\text{c}}/c$, where $v_{\text{LOS}}$ is the projection of the \ac{uav} velocity onto the \ac{los} direction. As shown in Fig.~\ref{fig:cfo_field_test}, during the on-track flight phase, when the \ac{uav} follows the predetermined trajectory for radio-map sampling, the theoretical \ac{cfo} and the estimated \ac{cfo} (centered) closely match. The larger discrepancies around some peaks and valleys are mainly observed during turning maneuvers, where the received signal becomes less stable. Overall, the strong agreement between the two curves confirms that the proposed framework can accurately track the \ac{cfo} from 5G \ac{nr} \ac{ss} burst sets and validates the effectiveness of the entire measurement system.

Fig.~\ref{fig:per_beam_radio_map} illustrates the per-beam radio maps in the $150\text{ m}$ low-altitude airspace. The coverage patterns
of the eight beams can be clearly distinguished: beam 0 provides broad-area coverage, while beams 1 to 7 form seven narrower beams that sweep from approximately north toward west. The received power does not exhibit a simple monotonic distance- or propagation-loss trend but instead shows irregular spatial variations, since the dominant energy in the low-altitude airspace is largely contributed by ground-reflected paths. For the same reason, different beams exhibit different power levels in the airspace, which mainly reflect variations in the underlying ground environment; for example, beams 1-3 appear weaker in regions with fewer reflective structures, whereas beams 4-7 are stronger where more reflectors are present. These observations highlight the intrinsic complexity of low-altitude coverage by terrestrial deployments.

Fig.~\ref{fig:per_cell_radio_map} illustrates the per-cell radio maps in the $150\text{ m}$ low-altitude airspace. In Fig.~\ref{fig:per_cell_radio_map}(a), the received-power map for representative cells is shown, where at each position the maximum power over all beams of that cell is taken. The dominant coverage regions of different cells can be clearly distinguished: the cell with \ac{pci} 45 mainly serves the north-west sector, the cell with \ac{pci} 47 covers the southern sector, and the cell with \ac{pci} 559 primarily illuminates the eastern side. To assess low-altitude communication performance, Fig.~\ref{fig:per_cell_radio_map}(b)
presents the corresponding \ac{sinr} radio map, partitioned into four ranges, \ac{sinr} $<-5\text{ dB}$, $[-5,0]\text{ dB}$, $[0,10]\text{ dB}$, and $>10\text{ dB}$, which can be interpreted as no coverage, low-rate, medium-rate, and high-rate regions, respectively, in line with typical 3GPP link-budget guidelines~\cite{3gpp:ts38521}. The results show that ground \acpl{bs} can provide wide-area low-rate coverage in the airspace via reflections, and medium-rate coverage around the main reflection regions. However, coverage holes appear in inter-cell areas where the reflected power is low and inter-cell interference is strong. Moreover, due to the large reflection loss and severe inter-cell interference (for example, between cells 45 and 47, whose ground beams are oriented towards different areas but whose reflected components overlap aloft) only very limited regions achieve high \ac{sinr}.
This observation highlights the inherent difficulty of sustaining high-rate links in interference-limited low-altitude deployments and
underscores the need for coordinated terrestrial \ac{bs} deployment that explicitly accounts for both ground and low-altitude users.

\section{Conclusion\label{sec:conclusion}}

This work proposed a multi-block \ac{cfo}-channel estimation algorithm for constructing per-cell per-beam low-altitude radio maps from synchronization signals. The \ac{crlb} analysis shows that the estimation error variance is determined by the aggregated \ac{sinr} across coherent blocks, and decreases with the number of usable blocks and with a smaller consumed kinematic budget. These results suggest that the proposed method is particularly well-suited to future multi-beam systems, while also providing useful guidelines on the platform kinematics required for accurate radio-map measurements. Simulations showed that the proposed method can detect and estimate synchronization signals at \ac{sinr} levels down to $-30~\text{dB}$. Field measurements with a \ac{uav}-mounted receiver confirmed that the proposed method can reliably obtain per-cell per-beam low-altitude radio maps and revealed that, despite strong
received power, the measured \ac{sinr} rarely exceeds $10~\text{dB}$. These results provide a fundamental basis for the design of aerial
communication systems and highlight the critical need for interference-aware design and optimization of terrestrial deployments to support future low-altitude aerial services.

\begin{figure}
\begin{centering}
\includegraphics[width=1\columnwidth]{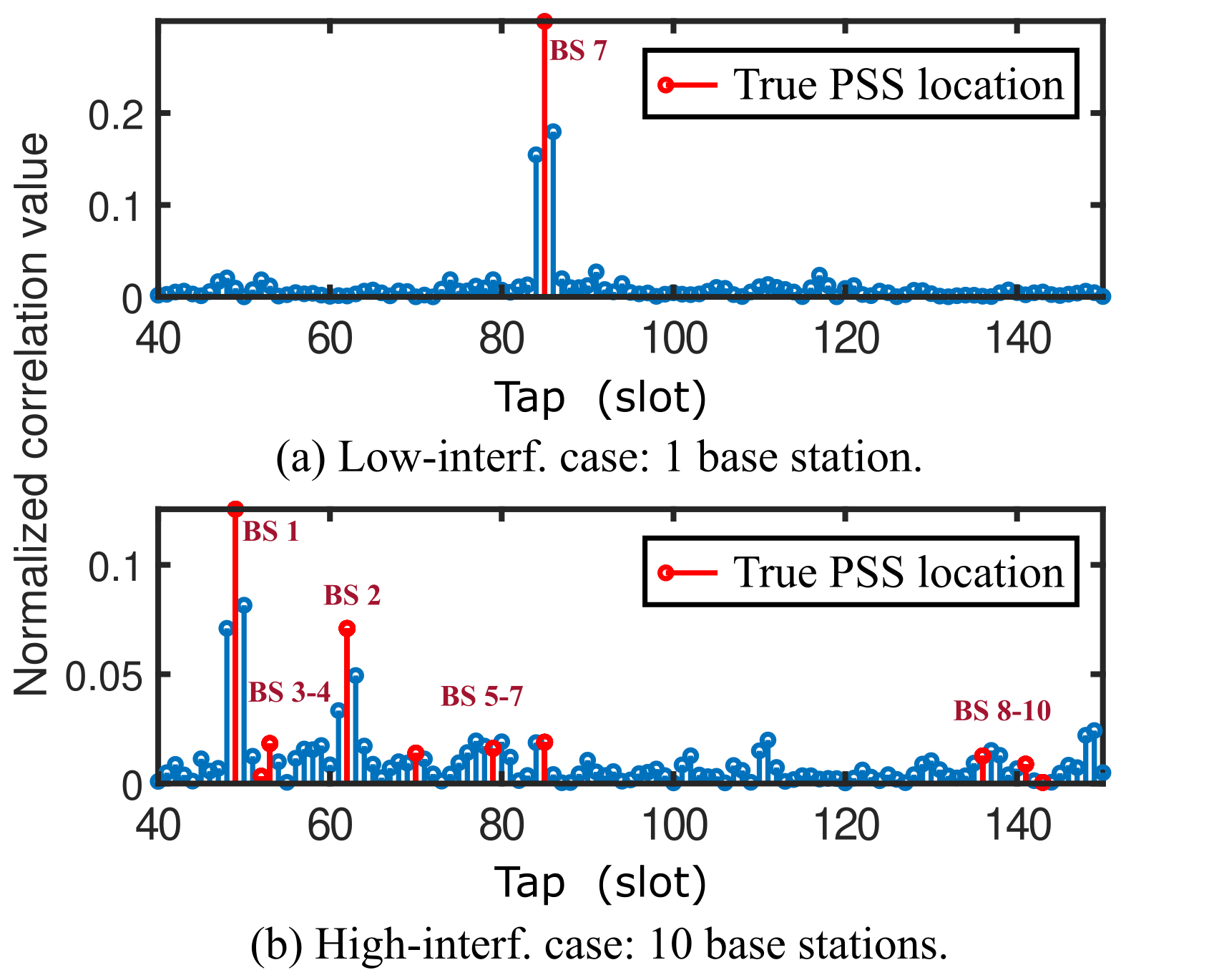}
\par\end{centering}
\caption{\label{fig:pss_estimation_error}Impact of interference on \ac{pss}-based
timing synchronization and detection. (a) In the low-interference
(single-cell) case, the correlation peak provides an unambiguous timing
estimate and \ac{pss} detection. (b) In the high-interference (10-cell)
case, the correlation profile is severely corrupted: strong peaks
from interfering cells create significant ambiguity, while the true
correlation peaks are either severely attenuated or completely suppressed.}
\end{figure}
\begin{figure}
\begin{centering}
\includegraphics[width=1\columnwidth]{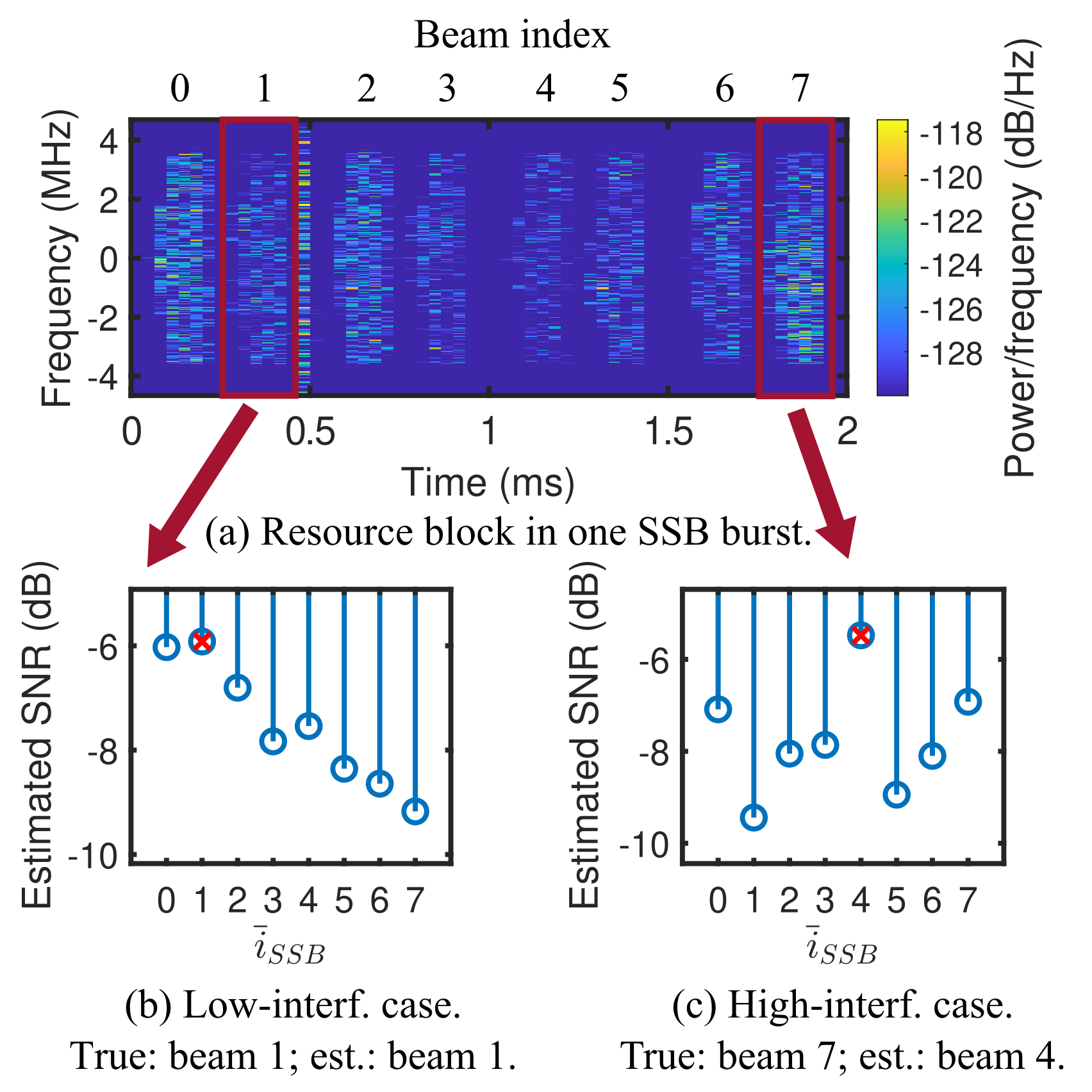}
\par\end{centering}
\caption{\label{fig:ssb_index_error}Impact of interference on \ac{dmrs}-based
\ac{ssb} index identification. (a) analyzes an empirically measured
\ac{ss} burst set. (b) For a low-interference beam, the estimated
\ac{snr} profile correctly identifies the true \ac{ssb} index (1).
(c) However, for a high-interference beam, co-channel interference
from other cells corrupts the \ac{snr} estimation, leading to the
erroneous identification of index 4 instead of the true index 7.}
\end{figure}

\appendices

\section{5G \ac{nr} Synchronization Framework, Detection Challenges, and
Proposed Pre-Processing \label{sec:5GNR}}

This Appendix introduces the 5G \ac{nr} \ac{ss} burst structure
and explains why it is suitable for per-cell, per-beam radio map construction
in low-altitude environments, then summarizes classical detection
and estimation pipelines, demonstrating their lack of robustness against
the severe interference typical of low-altitude environments. To overcome
these limitations, we finally present a new pre-processing algorithm
that leverages cross-burst properties to achieve robust initial detection.

\subsection{5G \ac{nr} \ac{ss} Burst Set Architecture}

The 5G \ac{nr} physical layer adopts a well-defined synchronization
and broadcast hierarchy for robust initial access. An \ac{ss}-burst
set is periodically transmitted with a typical periodicity of $20\text{ ms}$
and occupies at most $5\text{ ms}$, enabling beam sweeping via multiple
\acpl{ssb} on distinct transmit beams \cite{3gpp:ts38331}. The time-domain
mapping of each \ac{ssb} is dictated by the specific \ac{ssb} pattern
case (A\textendash E) and its index within the burst set \cite{3gpp:ts38213}.
For instance, under Case~C, up to eight \acpl{ssb} are time-multiplexed
within a brief window of approximately $2.5\text{ ms}$. Each individual
\ac{ssb} is transmitted via a distinct beamforming pattern, allowing
the \ac{bs} to sweep across different spatial directions for comprehensive
coverage.

Each \ac{ssb} comprises the \ac{pss}, \ac{sss}, and \ac{pbch}
with \ac{dmrs} \cite{3gpp:ts38211}. The \ac{pss} and \ac{sss}
sequences are deterministically derived from the \ac{pci} $N_{\text{ID}}^{\text{cell}}$:
\ac{pss} is a length-127 m-sequence uniquely determined by $N_{\text{ID}}^{(2)}=N_{\text{ID}}^{\text{cell}}\bmod3$;
the \ac{sss} is a length-127 m-sequence determined by $\big(N_{\text{ID}}^{(1)},N_{\text{ID}}^{(2)}\big)$,
where $N_{\text{ID}}^{(1)}=\left\lfloor N_{\text{ID}}^{\text{cell}}/3\right\rfloor $.
The position and content of \ac{pbch} \ac{dmrs} depends on the \ac{pci}
and \ac{ssb} index.

The structure of the \ac{ss} burst set enables low-altitude radio
mapping that is feasible, stable, and resilient to interference. Fine-grained
per-cell per-beam mapping follows from the deterministic embedding
of the \ac{pci} in the \ac{pss}/\ac{sss} and of the \ac{ssb} index
in the \ac{pbch} \ac{dmrs}. Stability and on-demand availability
result from the rapid and periodic \ac{ssb} transmissions. Moreover,
the \ac{ss} burst structure itself facilitates accurate detection
and estimation through (i) cross-block \ac{cfo} coherence over the
short burst duration ($\le5\text{ ms}$) and (ii) predetermined \ac{ssb}
positions and patterns, which are jointly exploited to achieve high-accuracy
processing under multi-cell interference.

\subsection{Conventional Detection Pipeline and Bottlenecks}

The conventional pipelines employ a sequential, multi-stage processing
chain to extract synchronization parameters and channel state information
\cite{TunRivGarMel:J23,ZhoCheYanChe:J24,YonSawNagSuy:J25,DonCheJuZho:J25,MathWorks:CellSearch}.
The process initiates with \ac{pss} detection in the time domain,
where the timing offset, \ac{cfo}, and partial cell identity ($N_{\text{ID}}^{(2)}$)
are jointly estimated by maximizing the correlation between the received
waveform and local \ac{pss} replicas. Following time-frequency compensation,
\ac{sss} detection is conducted in the frequency domain to resolve
the full \ac{pci} ($N_{\text{ID}}^{\text{cell}}$). Finally, \ac{pbch}
\ac{dmrs} processing identifies the specific beam index ($i_{\text{SSB}}$)
by matching the received signal in the frequency domain with the candidate
reference signals.

However, the above pipeline only works for the case of detecting the
strongest signal while treating all the interference as noise. As
illustrated in Fig.~\ref{fig:pss_estimation_error}, when the reception
contains signals from 10 \acpl{bs}, the \ac{pss} detection fails
for \acpl{bs} 3 to 10 although their \acpl{ss} are nearly orthogonal.
Likewise, the detection of the \ac{ssb} beam index is not possible
either under strong interference, as illustrated in Fig.~\ref{fig:ssb_index_error}.
Moreover, owing to the sequential nature of the conventional pipeline,
errors in time-frequency offset and root selection propagate downstream
and ultimately corrupt the estimated channel gains.

\subsection{Proposed Pre-Processing Algorithm}

\begin{figure}
\begin{centering}
\includegraphics[width=1\columnwidth]{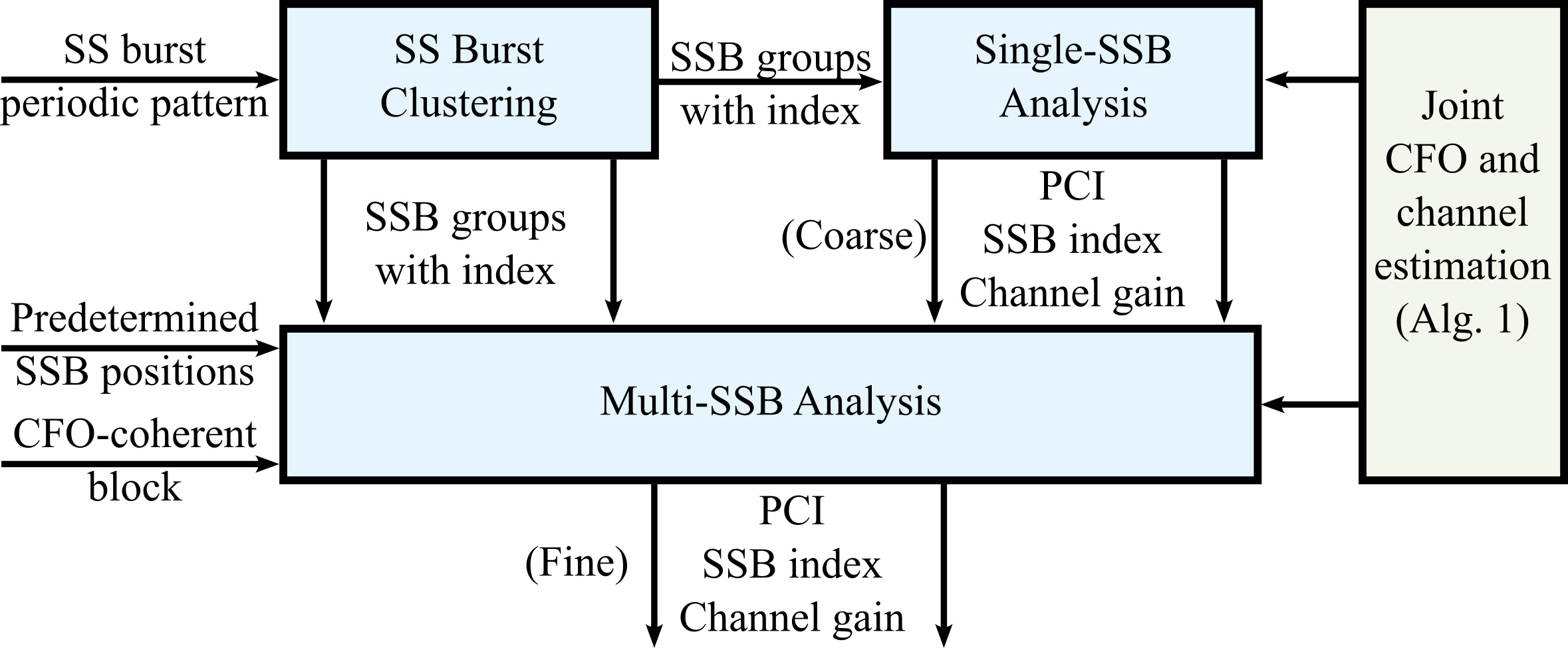}
\par\end{centering}
\caption{\label{fig:frame-work}Proposed cross-burst detection and estimation
framework. The algorithm follows a three-stage coarse-to-fine strategy,
including \ac{ss}-burst clustering, single-burst SIC-based analysis,
and cross-burst refinement, that exploits predetermined \ac{ssb}
positions and \ac{cfo} coherence to recover weak beams and produce
refined \ac{pci}, \ac{ssb}-index, and channel estimates.}
\end{figure}

We propose a cross-burst detection and estimation framework, illustrated
in Fig.~\ref{fig:frame-work}, that exploits the structured nature
of the \ac{ss} burst set. On the detection side, a non-cascaded procedure
is proposed that jointly processes multiple \acpl{ssb} to improve
detection efficiency and robustness in multi-cell interference (pre-processing).
On the estimation side, we leverage \ac{cfo} coherence within the
burst to construct multi-\ac{ssb} estimators, thereby enhancing the
accuracy of \ac{cfo} and channel-gain estimation, especially for
weak, interference-masked beams (Algorithm~\ref{alg:estimation}).
Next, we will introduce the proposed pre-processing algorithm.

\subsubsection{SSB Identification via Timing Information}

\begin{figure}
\begin{centering}
\includegraphics[width=1\columnwidth]{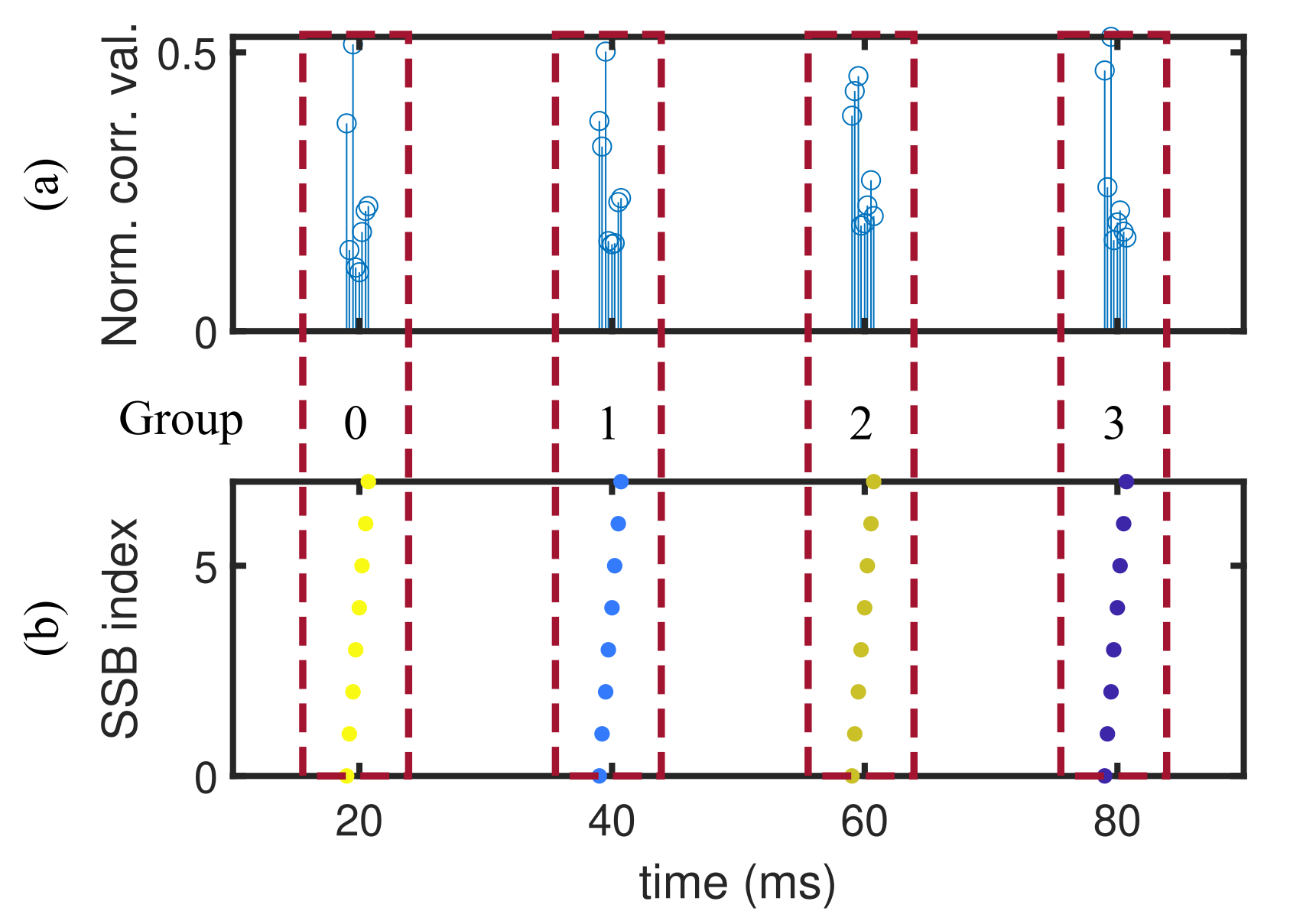}
\par\end{centering}
\caption{\label{fig:ssb_temperal_cluster}\ac{ssb} index identification via
temporal clustering of \ac{ss} bursts. The process consists of three
stages: 1) identifying candidate timing markers from the \ac{pss}
correlation envelope (a); 2) grouping these markers by their periodic
pattern into distinct clusters (colors in b); and 3) assigning an
\ac{ssb} index to each marker based on its relative timing within
its assigned cluster (vertical position in b).}
\end{figure}

We propose an \ac{ssb}-index identification method that exploits
the temporal regularity of the \ac{ss} burst set and does not rely
on prior \ac{pci} detection or highly accurate timing/frequency-offset
estimates. The key observation is that inter-\ac{bs} timing offsets
are much smaller than the spacing between \acpl{ssb}, since neighboring
\acpl{bs} are quasi-synchronous and the propagation-delay differences
over typical inter-site distances are negligible at the \ac{ssb}
time scale. Consequently, the signal received in the $i$th intra-\ac{ssb}
time window is dominated by the superposition of the $i$th \acpl{ssb}
from nearby \acpl{bs}, so each segment can be indexed by its position
within the burst pattern, independent of the \ac{pci}, as shown in
Fig.~\ref{fig:ssb_index_error}(a). In addition, the inter-burst
period ({\em e.g.}, $20\text{ ms}$ is much longer than the intra-burst
spacing (within $2.5\text{ ms}$), so standard time-domain clustering
can be used to group \acpl{ssb} by burst and then assign distinct
intra-burst indices.

To extract representative timing markers for each \ac{ssb}, we first
compute the \ac{pss}-based correlation envelope 
\[
\Lambda\left(t\right)\triangleq\max_{l\in\{0,1,2\}}\left|\frac{1}{N}\sum_{m=0}^{N-1}y\left[t+m\right]\left(c_{\text{PSS}}^{(l)}[t]\right)^{*}\right|^{2}
\]
where $c_{\text{PSS}}^{(l)}[t]$ presents the $l$th \ac{pss} time-domain
sequence. A structure-aware non-maximum suppression is then applied:
a local maximum $t_{x}$ is retained only if it is the unique dominant
peak within a guard window that is long enough to cover relative delays
across \acpl{bs}, yet short enough to exclude the \ac{pss} of adjacent
\acpl{ssb}. The surviving peaks serve as timing markers, which are
clustered according to the known \ac{ss}-burst periodicity and then
mapped to their relative intra-\ac{ssb} positions, yielding \ac{ssb}-index
assignments, as shown in Fig.~\ref{fig:ssb_temperal_cluster}, that
are \ac{pci}-agnostic and robust to residual timing and frequency
offsets.

\subsubsection{Complete Timing Offset Estimation and Full \ac{ssb} Detection}

Denote the set of estimated \ac{ssb} indices and timing offsets via
single-\ac{ssb} analysis for group $g$ and \ac{bs} $k$ by $\hat{\boldsymbol{\tau}}_{g,k}^{\text{single}}\triangleq\{\hat{\tau}_{g,k,p}^{\text{single}}\}_{p\in\mathcal{P}_{g,k}}$,
where $\hat{\tau}_{g,k,p}^{\text{single}}$ is the timing estimate
for \ac{ssb} $p$, and $\mathcal{P}_{g,k}\subseteq\mathcal{P}$ denotes
the index set of detected \acpl{ssb} in group $g$ for the $k$th
\ac{bs}. Then, each detected and estimated delay $\hat{\tau}_{g,k,p}^{\text{single}}$
is able to yield a candidate estimate of the absolute timing of \ac{ssb}
0 via 
\[
\hat{\tau}_{0}[p]=\hat{\tau}_{g,k,p}^{\text{single}}-\eta_{p},\,\forall p\in\mathcal{P}_{g,k}.
\]
Here, $\boldsymbol{\eta}=\{\eta_{p}\}_{p\in\mathcal{P}}$ with $\eta_{0}=0$
is the protocol-defined intra-\ac{ssb} timing pattern.

Stacking all candidates, the absolute timing of \ac{ssb} 0 can be
obtained by the least-squares estimate 
\[
\hat{\tau}_{0}^{\text{cross}}=\arg\min_{x}\frac{\sum_{p\in\mathcal{P}_{g,k}}\left|x-\hat{\tau}_{0}\left[p\right]\right|^{2}}{\left|\mathcal{P}_{g,k}\right|}=\frac{\sum_{p\in\mathcal{P}_{g,k}}\hat{\tau}_{0}\left[p\right]}{\left|\mathcal{P}_{g,k}\right|}.
\]
The resulting estimates for all \ac{ssb} positions in the $(g,k)$
burst are then given by the shifted protocol grid 
\begin{equation}
\hat{\boldsymbol{\tau}}_{g,k,p}^{\text{cross}}=\hat{\tau}_{0}^{\text{cross}}+\eta_{p},\,\forall p\in\mathcal{P}.\label{eq:cross_timing_offset_est}
\end{equation}

\section{Proof of Theorem \ref{thm:Kinematic_condition}\label{sec:proof_lem_cfo_drift}}

Let $\mathbf{p}\left(t\right)$ denote the position vector of the
receiver relative to the effective source, with the effective path
length denoted by $d\left(t\right)=||\mathbf{p}\left(t\right)||$.
The instantaneous \ac{doa} of the secular wavefront is given by the
unit vector $\mathbf{u}\left(t\right)=\mathbf{p}\left(t\right)/d\left(t\right)$.
The Doppler shift is determined by the projection of the receiver's
velocity vector $\mathbf{v}\left(t\right)$ onto $\mathbf{u}\left(t\right)$
\[
f_{d}\left(t\right)=\frac{1}{\lambda_{c}}\mathbf{v}\left(t\right)^{\text{T}}\mathbf{u}\left(t\right)
\]
where $\lambda_{c}$ is the carrier wavelength.

The maximum \ac{cfo} drift is bounded by the integral of the absolute
rate of change of the Doppler shift 
\[
\Delta f\le\int_{t_{0}}^{t_{0}+\delta_{t}}\left|\dot{f}_{d}\left(t\right)\right|dt\le\delta_{t}\cdot\max_{t\in\left[t_{0},t_{0}+\delta_{t}\right]}\left|\dot{f}_{d}\left(t\right)\right|.
\]
Differentiating the Doppler equation $f_{d}\left(t\right)$ with respect
to time yields two components 
\[
\dot{f}_{d}\left(t\right)=\frac{1}{\lambda_{\text{c}}}\left(\dot{\mathbf{v}}\left(t\right)^{\text{T}}\mathbf{u}\left(t\right)+\mathbf{v}\left(t\right)^{\text{T}}\dot{\mathbf{u}}\left(t\right)\right).
\]
Substituting the acceleration $\mathbf{a}\left(t\right)=\dot{\mathbf{v}}\left(t\right)$
and applying the Cauchy-Schwarz inequality, we have 
\[
\left|\dot{\mathbf{v}}\left(t\right)^{\text{T}}\mathbf{u}\left(t\right)\right|=\left|\mathbf{a}\left(t\right)^{\text{T}}\mathbf{u}\left(t\right)\right|\le\left\Vert \mathbf{a}\left(t\right)\right\Vert \left\Vert \mathbf{u}\left(t\right)\right\Vert \le\bar{a}.
\]
The time derivative of the direction unit vector $\mathbf{u}(t)=\mathbf{p}(t)/d(t)$
is given by 
\[
\dot{\mathbf{u}}\left(t\right)=\frac{\dot{\mathbf{p}}\left(t\right)}{d\left(t\right)}-\mathbf{p}\left(t\right)\frac{\dot{d}\left(t\right)}{d\left(t\right)^{2}}=\frac{1}{d\left(t\right)}\left(\mathbf{v}\left(t\right)-\mathbf{u}\left(t\right)\mathbf{u}\left(t\right)^{\text{T}}\mathbf{v}\left(t\right)\right).
\]
Denote $\mathbf{P}_{\bot}\left(t\right)\triangleq\mathbf{I}-\mathbf{u}\left(t\right)\mathbf{u}\left(t\right)^{\text{T}}$
as the orthogonal projector onto the hyperplane perpendicular to $\mathbf{u}\left(t\right)$.
Therefore
\begin{align*}
\left|\mathbf{v}\left(t\right)^{\text{T}}\dot{\mathbf{u}}\left(t\right)\right| & \le\left\Vert \mathbf{v}\left(t\right)\right\Vert \left\Vert \dot{\mathbf{u}}\left(t\right)\right\Vert =\left\Vert \mathbf{v}\left(t\right)\right\Vert \left\Vert \frac{1}{d\left(t\right)}\mathbf{P}_{\bot}\left(t\right)\mathbf{v}\left(t\right)\right\Vert \\
 & \le\frac{1}{d\left(t\right)}\left\Vert \mathbf{v}\left(t\right)\right\Vert ^{2}\sin\theta\left(t\right)\le\frac{\bar{v}^{2}\bar{s}}{\underline{d}}.
\end{align*}
Substituting these bounds back into the Doppler shift yields 
\[
\Delta f\le\frac{\delta_{t}}{\lambda_{\text{c}}}\left(\bar{a}+\frac{\bar{v}^{2}\bar{s}}{\underline{d}}\right).
\]
To make sure $\Delta f\le\bar{\delta}_{f}$, it suffices to require
\[
\bar{a}+\frac{\bar{v}^{2}\bar{s}}{\underline{d}}\le\frac{\lambda_{\text{c}}\bar{\delta}_{f}}{\delta_{t}}.
\]

\section{Proof of Lemma \ref{lem:x_r_yk}\label{sec:proof_lem_x_r_yk}}

Start from the first synchronization sequence, that is, $i=0$, according
to the definition of $y[m]$ in (\ref{eq:def_rx_signal}), the cross-correlation
can be expressed as 
\begin{align*}
 & r_{y,k}^{p,0}\\
 & =\frac{1}{N}\sum_{m=\tau_{k}}^{\tau_{k}+N-1}\left(\sum_{q=1}^{K}\alpha_{q}^{\left(p\right)}c_{q,0}\left[m-\tau_{q}\right]e^{-j\omega_{q}m}+\nu\left[m\right]\right)\\
 & \quad\quad\quad\quad\quad\quad\quad\quad\quad\quad\quad\quad\quad\quad\quad\quad\times c_{k,0}^{\text{H}}\left[m-\tau_{k}\right]
\end{align*}
Decompose the sum into: i) desired signal $q=k$, ii) multi-\ac{bs}
interference $q\neq k$, and iii) noise.

For $q=k$, the cross-correlation for desired signal is 
\begin{multline*}
\frac{1}{N}\sum_{m=\tau_{k}}^{\tau_{k}+N-1}\alpha_{k}^{\left(p\right)}c_{k,0}\left[m-\tau_{k}\right]e^{-j\omega_{k}m}c_{k,0}^{\text{H}}\left[m-\tau_{k}\right]\\
=\alpha_{k}^{\left(p\right)}r_{k,k}\left(0,\omega_{k}\right)=\alpha_{k}^{\left(p\right)}r_{k}\left(\omega_{k}\right).
\end{multline*}
For $q\neq k$, the cross-correlation for interference signal is 
\begin{multline*}
\sum_{q\neq k}\frac{1}{N}\sum_{m=\tau_{k}}^{\tau_{k}+N-1}\alpha_{q}^{\left(p\right)}c_{q,0}\left[m-\tau_{q}\right]e^{-j\omega_{q}m}c_{k,0}^{\text{H}}\left[m-\tau_{k}\right]\\
=\sum_{q\neq k}\alpha_{q}^{\left(p\right)}r_{k,q}\left(\tau_{k}-\tau_{q},\omega_{q}\right)=\mathcal{CN}\left(0,\sum_{q\neq k}\alpha_{q}^{\left(p\right)2}\sigma_{c}^{2}/N\right).
\end{multline*}
For the noise, the cross-correlation is 
\begin{multline*}
\frac{1}{N}\sum_{m=\tau_{k}}^{\tau_{k}+N-1}\nu\left[m\right]c_{k,0}^{\text{H}}\left[m-\tau_{k}\right]\\
=\mathcal{CN}\left(0,\frac{1}{N}\sum_{m=0}^{N-1}\left|c_{k,0}\left[m\right]\right|^{2}\sigma_{n}^{2}\right)=\mathcal{CN}\left(0,\sigma_{n}^{2}/N\right).
\end{multline*}
Putting i) to iii) together gives 
\[
r_{y,k}^{p,0}=\alpha_{k}^{\left(p\right)}r_{k}\left(\omega_{k}\right)+\mathcal{CN}\left(0,\sigma_{k,p,0}^{2}\right)
\]
where $\sigma_{k,p,0}^{2}=(\sum_{q\neq k}|\alpha_{q}^{\left(p\right)}\sigma_{c}|^{2}+\sigma_{n}^{2})/N$.

Similarly, consider $i=1$, we have 
\[
r_{y,k}^{p,1}=\mu_{k}\alpha_{k}^{\left(p\right)}r_{k}\left(\omega_{k}\right)e^{-j\omega_{k}\tau_{\text{c}}}+\mathcal{CN}\left(0,\sigma_{k,p,1}^{2}\right)
\]
where $\sigma_{k,p,1}^{2}=(\sum_{q\neq k}|\mu_{q}\alpha_{q}^{\left(p\right)}\sigma_{c}|^{2}+\sigma_{n}^{2})/N$.

In summary, the cross-correlation $r_{y,k}^{p,i}$ can be expressed
as 
\[
r_{y,k}^{p,i}=\alpha_{k}^{\left(p\right)}r_{k}\left(\omega_{k}\right)\left(\mu_{k}e^{-j\omega_{k}\tau_{\text{c}}}\right)^{i}+\mathcal{CN}\left(0,\sigma_{k,p,i}^{2}\right)
\]
with $\sigma_{k,p,i}^{2}=(\sum_{q\neq k}|\mu_{q}^{i}\alpha_{q}^{\left(p\right)}\sigma_{c}|^{2}+\sigma_{n}^{2})/N$.

\section{Proof of Proposition \ref{prop:joint_est_omega}\label{sec:proof_prop_joint_est}}

Since the autocorrelation function $r_{k}\left(\omega_{k}\right)$
depends on the \ac{cfo} variable $\omega_{k}$, we introduce an
auxiliary variable $\gamma_{k}\triangleq r_{k}\left(\omega_{k}\right)$
to absorbs and relaxes the \ac{cfo} dependence. Consequently, the
unknown parameters now consist of $\boldsymbol{\theta}\triangleq[\omega_{k},\gamma_{k},\gamma_{k}^{\text{H}}]^{\text{T}}$,where
$\omega_{k}$ is real and $(\gamma_{k},\gamma_{k}^{\text{H}})$ is
complex. Using this reparametrization, the log-likelihood function
(\ref{eq:def_f}) becomes 
\begin{align}
f\left(\boldsymbol{\theta}\right) & =\sum_{p\in\mathcal{P}}-\frac{\left|r_{y,k}^{p,0}-\alpha_{k}^{(p)}\gamma_{k}\right|^{2}}{\sigma_{k,p,0}^{2}}\nonumber \\
 & \quad+\sum_{p\in\mathcal{P}}-\frac{\left|r_{y,k}^{p,1}-\alpha_{k}^{(p)}\gamma_{k}\mu_{k}e^{-j\omega_{k}\tau_{\text{c}}}\right|^{2}}{\sigma_{k,p,1}^{2}}.\label{eq:reparametrization}
\end{align}

\subsubsection{Relaxed \ac{ml} estimator}

Denote the cumulative average \ac{sinr} for each reference signal
as 
\begin{equation}
\Gamma_{0}=\sum_{p\in\mathcal{P}}\frac{\left|\alpha_{k}^{(p)}\right|^{2}}{\sigma_{k,p,0}^{2}},\quad\Gamma_{1}=\sum_{p\in\mathcal{P}}\frac{\left|\mu_{k}\alpha_{k}^{(p)}\right|^{2}}{\sigma_{k,p,1}^{2}}\label{eq:def_gamma}
\end{equation}
and the cumulative average \ac{sinr} realization for each reference
signal as 
\begin{equation}
\psi_{0}\triangleq\sum_{p\in\mathcal{P}}\frac{\alpha_{k}^{(p)}\left(r_{y,k}^{p,0}\right)^{\text{H}}}{\sigma_{k,p,0}^{2}},\quad\psi_{1}\triangleq\sum_{p\in\mathcal{P}}\frac{\alpha_{k}^{(p)}\mu_{k}\left(r_{y,k}^{p,1}\right)^{\text{H}}}{\sigma_{k,p,1}^{2}}.\label{eq:def_phi}
\end{equation}
Then, taking differentiation for $(\omega_{k},\gamma_{k},\gamma_{k}^{\text{H}})$,
we have 
\begin{equation}
\frac{\partial f}{\partial\omega_{k}}=2\tau_{\text{c}}\text{Im}\left\{ \psi_{1}\gamma_{k}e^{-j\omega_{k}\tau_{\text{c}}}\right\} \label{eq:pf_pw-1}
\end{equation}
\begin{equation}
\frac{\partial f}{\partial\gamma_{k}}=\psi_{0}+\psi_{1}e^{-j\omega_{k}\tau_{\text{c}}}-\left(\Gamma_{0}+\Gamma_{1}\right)\gamma_{k}^{\text{H}}\label{eq:pf_pr}
\end{equation}
\begin{equation}
\frac{\partial f}{\partial\gamma_{k}^{\text{H}}}=\psi_{0}^{\text{H}}+\psi_{1}^{\text{H}}e^{j\omega_{k}\tau_{\text{c}}}-\left(\Gamma_{0}+\Gamma_{1}\right)\gamma_{k}.\label{eq:pf_prh}
\end{equation}

Taking $\frac{\partial f}{\partial\omega_{k}}=0$, we have 
\begin{equation}
\omega_{k}\tau_{\text{c}}=\angle\left\{ \psi_{1}\gamma_{k}\right\} \label{eq:omega_r}
\end{equation}
and taking $\frac{\partial f}{\partial\gamma_{k}^{\text{H}}}=0$,
we have 
\begin{equation}
\gamma_{k}=\frac{\psi_{0}^{\text{H}}+\psi_{1}^{\text{H}}e^{j\omega_{k}\tau_{\text{c}}}}{\left(\Gamma_{0}+\Gamma_{1}\right)}.\label{eq:r}
\end{equation}
Substituting (\ref{eq:r}) into (\ref{eq:omega_r}), we have 
\[
\omega_{k}\tau_{\text{c}}=\angle\left\{ \frac{\psi_{1}\psi_{0}^{\text{H}}+\left|\psi_{1}^{\text{H}}\right|^{2}e^{j\omega_{k}\tau_{\text{c}}}}{\left(\Gamma_{0}+\Gamma_{1}\right)}\right\} \stackrel{(a)}{=}\angle\left\{ \frac{\psi_{1}\psi_{0}^{\text{H}}}{\left|\psi_{1}^{\text{H}}\right|^{2}}+e^{j\omega_{k}\tau_{\text{c}}}\right\} 
\]
where (a) holds because $\angle\left\{ \alpha x\right\} =\angle\left\{ x\right\} $
for any non-zero real $\alpha$ and here we set $\alpha=(\Gamma_{0}+\Gamma_{1})/|\psi_{1}^{\text{H}}|^{2}$.
Then, we have 
\begin{align*}
 & \angle\left\{ \frac{\psi_{1}\psi_{0}^{\text{H}}}{\left|\psi_{1}^{\text{H}}\right|^{2}}+e^{j\omega_{k}\tau_{\text{c}}}\right\} -\omega_{k}\tau_{\text{c}}\\
 & =\angle\left\{ \left(\frac{\psi_{1}\psi_{0}^{\text{H}}}{\left|\psi_{1}^{\text{H}}\right|^{2}}e^{-j\omega_{k}\tau_{c}}+1\right)\right\} =0
\end{align*}
which means $\angle\left\{ \left(\frac{\psi_{1}\psi_{0}^{\text{H}}}{\left|\psi_{1}^{\text{H}}\right|^{2}}e^{-j\omega_{k}\tau_{c}}\right)\right\} =0$
then 
\[
\omega_{k}\tau_{c}=\angle\left\{ \frac{\psi_{1}\psi_{0}^{\text{H}}}{\left|\psi_{1}^{\text{H}}\right|^{2}}\right\} =\angle\left\{ \psi_{1}\psi_{0}^{\text{H}}\right\} .
\]
As a result, the \ac{ml} estimator is 
\begin{equation}
\omega_{k}=\angle\left\{ \psi_{1}\psi_{0}^{\text{H}}\right\} /\tau_{c}.\label{eq:est_omega_multi_A}
\end{equation}

\subsubsection{Equivalence to Constrained \ac{ml}}

The relaxed estimator coincides with the constrained \ac{ml} estimator
if the relaxed optimum satisfies the original constraint $\gamma_{k}=r_{k}(\omega_{k})$.

From the relaxed solution, substituting $\omega_{k}$ in (\ref{eq:est_omega_multi_A})
back into $\gamma_{k}$ in (\ref{eq:r}) gives 
\[
\gamma_{k}=\frac{\psi_{0}^{\text{H}}+\psi_{1}^{\text{H}}e^{j\angle\left\{ \psi_{1}\psi_{0}^{\text{H}}\right\} }}{\left(\Gamma_{0}+\Gamma_{1}\right)}=\frac{|\psi_{0}|+\left|\psi_{1}\right|}{\Gamma_{0}+\Gamma_{1}}e^{-j\angle\psi_{0}}.
\]
Therefore, the relaxed estimator is equivalent to the constrained
\ac{ml} estimator if
\begin{align*}
r_{k}\left(\omega_{k}\right) & =\frac{|\psi_{0}|+\left|\psi_{1}\right|}{\Gamma_{0}+\Gamma_{1}}e^{-j\angle\psi_{0}}.
\end{align*}

\section{Proof of Proposition \ref{prop:crlb}\label{sec:proof_prop_crlb}}

The cumulative correlations can be expressed as $\psi_{i}=\Gamma_{i}\gamma_{k}^{\text{H}}e^{j\omega_{k}\tau_{c}i}+\eta_{i}$
with $\eta_{i}\sim\mathcal{CN}(0,\Gamma_{i})$, where $i\in\{0,1\}$.
Normalizing by $\Gamma_{i}\gamma_{k}^{\text{H}}e^{j\omega_{k}\tau_{c}i}$
gives $\psi_{i}/(\Gamma_{i}\gamma_{k}^{\text{H}}e^{j\omega_{k}\tau_{c}i})=1+\epsilon_{i}$
with $\epsilon_{i}\sim\mathcal{CN}(0,1/(\Gamma_{i}|\gamma_{k}|^{2})$.
Then, we have 
\[
\hat{\omega}_{k}\tau_{c}=\angle\left\{ \psi_{1}\psi_{0}^{\text{H}}\right\} =\angle\left\{ |\gamma_{k}|^{2}e^{j\omega_{k}\tau_{c}}\left(1+\epsilon_{1}\right)\left(1+\epsilon_{0}\right)^{\text{H}}\right\} .
\]
Therefore,
\[
\hat{\omega}_{k}-\omega_{k}=\frac{1}{\tau_{c}}\angle\left\{ \left(1+\epsilon_{1}\right)\left(1+\epsilon_{0}\right)^{\text{H}}\right\} .
\]
 By the first-order Taylor expansion around $0+j0$, 
\[
\hat{\omega}_{k}-\omega_{k}=\frac{1}{\tau_{c}}\left(\text{Im}\left\{ \epsilon_{1}\right\} -\text{Im}\left\{ \epsilon_{0}\right\} +O_{p}\left(\left|\epsilon_{0}\right|^{2}+\left|\epsilon_{1}\right|^{2}\right)\right),
\]
where $O_{p}$ denotes stochastic order, \emph{i.e.}, $X_{n}=O_{p}\left(a_{n}\right)$
means $X_{n}/a_{n}$ is bounded \cite{Wol:B07}.

Since $\text{Im}\left\{ \epsilon_{i}\right\} \sim\mathcal{N}(0,1/(2\Gamma_{i}|\gamma_{k}|^{2})$
and $\left|\epsilon_{i}\right|^{2}=O_{p}(\mathbb{E}\{|\epsilon_{i}|^{2}\})$
based on Chebyshev's inequality, then $\hat{\omega}_{k}-\omega_{k}\overset{a}{\sim}\mathcal{N}(0,(\Gamma_{0}^{-1}+\Gamma_{1}^{-1})/(2\tau_{c}^{2}|\gamma_{k}|^{2}))$
with the approximation remainder $O_{p}((\Gamma_{0}^{-1}+\Gamma_{1}^{-1})/(\tau_{c}|\gamma_{k}|^{2}))$.

The same asymptotic variance (\ac{crlb}) can also be obtained from
\ac{fim} by applying the standard asymptotic property of the \ac{ml}
estimator \cite{Kay:B93}.

\bibliographystyle{IEEEtran}
\bibliography{IEEEabrv,bib_files/StringDefinitions,bib_files/BL}

\begin{thebibliography}{10}
\providecommand{\url}[1]{#1}
\csname url@samestyle\endcsname
\providecommand{\newblock}{\relax}
\providecommand{\bibinfo}[2]{#2}
\providecommand{\BIBentrySTDinterwordspacing}{\spaceskip=0pt\relax}
\providecommand{\BIBentryALTinterwordstretchfactor}{4}
\providecommand{\BIBentryALTinterwordspacing}{\spaceskip=\fontdimen2\font plus
\BIBentryALTinterwordstretchfactor\fontdimen3\font minus
  \fontdimen4\font\relax}
\providecommand{\BIBforeignlanguage}[2]{{%
\expandafter\ifx\csname l@#1\endcsname\relax
\typeout{** WARNING: IEEEtran.bst: No hyphenation pattern has been}%
\typeout{** loaded for the language `#1'. Using the pattern for}%
\typeout{** the default language instead.}%
\else
\language=\csname l@#1\endcsname
\fi
#2}}
\providecommand{\BIBdecl}{\relax}
\BIBdecl

\bibitem{SonLinWanSun:M25}
M.~Song, Y.~Lin, J.~Wang, G.~Sun, C.~Dong, N.~Ma, D.~Niyato, and P.~Zhang,
  ``Trustworthy intelligent networks for low-altitude economy,'' \emph{{IEEE}
  Commun. Mag.}, vol.~63, no.~7, pp. 72--79, 2025.

\bibitem{JiaLiZhuLi:M25}
Y.~Jiang, X.~Li, G.~Zhu, H.~Li, J.~Deng, K.~Han, C.~Shen, Q.~Shi, and R.~Zhang,
  ``Integrated sensing and communication for low altitude economy:
  Opportunities and challenges,'' \emph{{IEEE} Commun. Mag.}, pp. 1--7, 2025.

\bibitem{YanDinMaoLin:J19}
J.~Yang, M.~Ding, G.~Mao, Z.~Lin, D.-G. Zhang, and T.~H. Luan, ``Optimal base
  station antenna downtilt in downlink cellular networks,'' \emph{IEEE Trans.
  on Wireless Commun.}, vol.~18, no.~3, pp. 1779--1791, 2019.

\bibitem{MdIsaWalAru:J21}
M.~U.~C. Md, G.~Ismail, S.~Walid, and B.~Arupjyoti, ``Ensuring reliable
  connectivity to cellular-connected {UAVs} with up-tilted antennas and
  interference coordination,'' \emph{ITU Journal on Future and Evolving
  Technologies}, vol.~2, no.~2, pp. 165--185, 2021.

\bibitem{GerGarAzaLoz:J24}
G.~Geraci, A.~Garcia-Rodriguez, M.~M. Azari, A.~Lozano, M.~Mezzavilla,
  S.~Chatzinotas, Y.~Chen, S.~Rangan, and M.~D. Renzo, ``What will the future
  of {UAV} cellular communications be? a flight from {5G} to {6G},''
  \emph{{IEEE} Commun. Surveys Tuts.}, vol.~24, no.~3, pp. 1304--1335, 2022.

\bibitem{CheLiSunCui:M26}
J.~Chen, B.~Li, H.~Sun, S.~Cui, and N.~Pappas, ``Predictive communications for
  low-altitude networks,'' \emph{IEEE Internet Things Mag.}, pp. 1--8, 2026.

\bibitem{ChaLau:J22}
S.~Chai and V.~K. Lau, ``Mixed-timescale request-driven user association,
  trajectory and radio resource control for cache-enabled multi-{UAV}
  networks,'' \emph{{IEEE} Trans. Signal Process.}, vol.~70, pp. 4997--5011,
  2022.

\bibitem{ZhoZenLiYan:A25}
\BIBentryALTinterwordspacing
Z.~Zhou, Y.~Zeng, C.~Li, F.~Yang, Y.~Chen, and J.~Joung, ``Full-dimensional
  beamforming for multi-user {MIMO-OFDM} {ISAC} for low-altitude {UAV} with
  zero sensing resource allocation,'' 2025. [Online]. Available:
  \url{https://arxiv.org/abs/2508.06428}
\BIBentrySTDinterwordspacing

\bibitem{RomKim:M22}
D.~Romero and S.-J. Kim, ``Radio map estimation: A data-driven approach to
  spectrum cartography,'' \emph{{IEEE} Signal Process. Mag.}, vol.~39, no.~6,
  pp. 53--72, 2022.

\bibitem{Sunche:J24}
H.~Sun and J.~Chen, ``Integrated interpolation and block-term tensor
  decomposition for spectrum map construction,'' \emph{{IEEE} Trans. Signal
  Process.}, vol.~72, pp. 3896--3911, 2024.

\bibitem{ShrFuHon:J22}
S.~Shrestha, X.~Fu, and M.~Hong, ``Deep spectrum cartography: Completing radio
  map tensors using learned neural models,'' \emph{{IEEE} Trans. Signal
  Process.}, vol.~70, pp. 1170--1184, 2022.

\bibitem{WanZhaNieYu:M25}
H.~Wang, J.~Zhang, G.~Nie, L.~Yu, Z.~Yuan, T.~Li, J.~Wang, and G.~Liu,
  ``Digital twin channel for {6G}: Concepts, architectures and potential
  applications,'' \emph{{IEEE} Commun. Mag.}, vol.~63, no.~3, pp. 24--30, 2025.

\bibitem{HeAiGuaWan:J19}
D.~He, B.~Ai, K.~Guan, L.~Wang, Z.~Zhong, and T.~K\"{u}rner, ``The design and
  applications of high-performance ray-tracing simulation platform for {5G} and
  beyond wireless communications: A tutorial,'' \emph{{IEEE} Commun. Surveys
  Tuts.}, vol.~21, no.~1, pp. 10--27, 2019.

\bibitem{3gpp:ts38331}
{3rd Generation Partnership Project (3GPP)}, ``{NR; Radio Resource Control
  (RRC); Protocol specification},'' 3GPP, Technical Specification (TS) 38.331,
  Jul 2025, release 18.

\bibitem{3gpp:ts38213}
------, ``{NR; Physical layer procedures for control},'' 3GPP, Technical
  Specification (TS) 38.213, Jul 2025, release 18.

\bibitem{3gpp:ts38211}
------, ``{NR; Physical channels and modulation},'' 3GPP, Technical
  Specification (TS) 38.211, Apr 2025, release 18.

\bibitem{TunRivGarMel:J23}
R.~Tuninato, D.~G. Riviello, R.~Garello, B.~Melis, and R.~Fantini, ``A
  comprehensive study on the synchronization procedure in {5G} {NR} with
  {3GPP}-compliant link-level simulator,'' \emph{{EURASIP} J. Wireless Com.
  Network}, vol. 2023, no.~1, p. 111, 2023.

\bibitem{ZhoCheYanChe:J24}
X.~Zhou, L.~Chen, Y.~Ruan, and R.~Chen, ``Indoor localization with multi-beam
  of {5G} new radio signals,'' \emph{IEEE Trans. on Wireless Commun.}, vol.~23,
  no.~9, pp. 11\,260--11\,275, 2024.

\bibitem{YonSawNagSuy:J25}
S.~Yoneda, M.~Sawahashi, S.~Nagata, and S.~Suyama, ``Prach transmission
  employing carrier frequency offset pre-compensation based on measurement at
  {UE} for {NR} uplink,'' \emph{IEICE Trans. on Commun.}, vol. E108-B, no.~3,
  pp. 330--338, 2025.

\bibitem{DonCheJuZho:J25}
W.~Dong, L.~Chen, Z.~Ju, T.~Zhou, Z.~Liu, and R.~Chen, ``Beam-switching-based
  time-of-arrival ranging on commercial {5G} {NR} signals for outdoor
  positioning,'' \emph{IEEE Trans. on Wireless Commun.}, pp. 1--1, 2025.

\bibitem{MathWorks:CellSearch}
{MathWorks}, ``{NR} cell search and {MIB} and {SIB1} recovery,''
  \url{https://se.mathworks.com/help/5g/ug/nr-cell-search-and-mib-and-sib1-recovery.html},
  2025, accessed: 2025-11-03.

\bibitem{MorKuoPun:J07}
M.~Morelli, C.-C.~J. Kuo, and M.-O. Pun, ``Synchronization techniques for
  orthogonal frequency division multiple access ({OFDMA}): A tutorial review,''
  \emph{Proc. {IEEE}}, vol.~95, no.~7, pp. 1394--1427, 2007.

\bibitem{ParChu:J12}
J.~Park and J.~Chun, ``Improved lattice reduction-aided {MIMO} successive
  interference cancellation under imperfect channel estimation,'' \emph{{IEEE}
  Trans. Signal Process.}, vol.~60, no.~6, pp. 3346--3351, 2012.

\bibitem{VanSanBor:J97}
J.~van~de Beek, M.~Sandell, and P.~Borjesson, ``{ML} estimation of time and
  frequency offset in {OFDM} systems,'' \emph{{IEEE} Trans. Signal Process.},
  vol.~45, no.~7, pp. 1800--1805, 1997.

\bibitem{LvHuaJie:J05}
T.~Lv, H.~Li, and J.~Chen, ``Joint estimation of symbol timing and carrier
  frequency offset of ofdm signals over fast time-varying multipath channels,''
  \emph{{IEEE} Trans. Signal Process.}, vol.~53, no.~12, pp. 4526--4535, 2005.

\bibitem{NasMehBloDur:J12}
A.~A. Nasir, H.~Mehrpouyan, S.~D. Blostein, S.~Durrani, and R.~A. Kennedy,
  ``Timing and carrier synchronization with channel estimation in multi-relay
  cooperative networks,'' \emph{{IEEE} Trans. Signal Process.}, vol.~60, no.~2,
  pp. 793--811, 2012.

\bibitem{GesHanHuaSha:J10}
D.~Gesbert, S.~Hanly, H.~Huang, S.~Shamai~Shitz, O.~Simeone, and W.~Yu,
  ``Multi-cell {MIMO} cooperative networks: A new look at interference,''
  \emph{{IEEE} J. Sel. Areas Commun.}, vol.~28, no.~9, pp. 1380--1408, 2010.

\bibitem{XuLarJorLi:J25}
Y.~Xu, E.~G. Larsson, E.~A. Jorswieck, X.~Li, S.~Jin, and T.-H. Chang,
  ``Distributed signal processing for extremely large-scale antenna array
  systems: State-of-the-art and future directions,'' \emph{{IEEE} Trans. Signal
  Process.}, vol.~19, no.~2, pp. 304--330, 2025.

\bibitem{ZarCav:J08}
B.~W. Zarikoff and J.~K. Cavers, ``Multiple frequency offset estimation for the
  downlink of coordinated {MIMO} systems,'' \emph{{IEEE} J. Sel. Areas
  Commun.}, vol.~26, no.~6, pp. 901--912, 2008.

\bibitem{WanXiaYin:J08}
H.~Wang, X.-G. Xia, and Q.~Yin, ``Computationally efficient equalization for
  asynchronous cooperative communications with multiple frequency offsets,''
  \emph{IEEE Trans. on Wireless Commun.}, vol.~8, no.~2, pp. 648--655, 2009.

\bibitem{TsaHuaCheYan:J13}
Y.-R. Tsai, H.-Y. Huang, Y.-C. Chen, and K.-J. Yang, ``Simultaneous multiple
  carrier frequency offsets estimation for coordinated multi-point transmission
  in {OFDM} systems,'' \emph{IEEE Trans. on Wireless Commun.}, vol.~12, no.~9,
  pp. 4558--4568, 2013.

\bibitem{SalNasMehXia:J17}
O.~H. Salim, A.~A. Nasir, H.~Mehrpouyan, and W.~Xiang, ``Multi-relay
  communications in the presence of phase noise and carrier frequency
  offsets,'' \emph{IEEE Trans. on Commun.}, vol.~65, no.~1, pp. 79--94, 2017.

\bibitem{ZhaGaoJinLin:J18}
W.~Zhang, F.~Gao, S.~Jin, and H.~Lin, ``Frequency synchronization for uplink
  massive mimo systems,'' \emph{IEEE Trans. on Wireless Commun.}, vol.~17,
  no.~1, pp. 235--249, 2018.

\bibitem{KunUnnSarLar:J24}
U.~Kunnath~Ganesan, R.~Sarvendranath, and E.~G. Larsson, ``{BeamSync}:
  Over-the-air synchronization for distributed massive {MIMO} systems,''
  \emph{IEEE Trans. on Wireless Commun.}, vol.~23, no.~7, pp. 6824--6837, 2024.

\bibitem{SinKumMajSat:J25}
S.~Singh, S.~Kumar, S.~Majhi, U.~Satija, and C.~Yuen, ``Blind carrier frequency
  offset estimation techniques for next-generation multicarrier communication
  systems: Challenges, comparative analysis, and future prospects,''
  \emph{{IEEE} Commun. Surveys Tuts.}, vol.~27, no.~1, pp. 1--36, 2025.

\bibitem{TseVis:B05}
D.~Tse and P.~Viswanath, \emph{Fundamentals of wireless communication}.\hskip
  1em plus 0.5em minus 0.4em\relax Cambridge university press, 2005.

\bibitem{Cla:J68}
R.~H. Clarke, ``A statistical theory of mobile-radio reception,'' \emph{The
  Bell System Technical Journal}, vol.~47, no.~6, pp. 957--1000, 1968.

\bibitem{WilCox:B94}
W.~C. Jakes and D.~C. Cox, \emph{Microwave mobile communications}.\hskip 1em
  plus 0.5em minus 0.4em\relax New York, NY, USA: IEEE Press / Wiley, 1994.

\bibitem{3gpp:ts38521}
{3rd Generation Partnership Project (3GPP)}, ``{5G; NR; User Equipment (UE)
  conformance specification; Radio transmission and reception; Part 1: Range 1
  standalone},'' 3GPP, Technical Specification (TS) 38.521-1, Feb 2025, 3GPP TS
  38.521-1 version 18.5.0 Release 18.

\bibitem{Wol:B07}
K.~M. Wolter and K.~M. Wolter, \emph{Introduction to variance
  estimation}.\hskip 1em plus 0.5em minus 0.4em\relax Springer, 2007, vol.~53.

\bibitem{Kay:B93}
S.~M. Kay, \emph{Fundamentals of statistical signal processing, Volume I:
  estimation theory}.\hskip 1em plus 0.5em minus 0.4em\relax Prentice-Hall,
  Inc., 1993.

\end{thebibliography}

\end{document}